\documentclass[12pt]{elsarticle}

\journal{Nucl. Instrum. Methods A}

\usepackage[margin={2cm}]{geometry}
\usepackage{rotating}
\usepackage{multirow}
\usepackage{amsmath}    
\usepackage{xspace}     
\usepackage{units}      
\usepackage{hyperref}   
\hypersetup{}
\usepackage[english]{babel} 
\usepackage{color}
\usepackage{multicol}
\usepackage{threeparttable}
\usepackage{refcount}
\usepackage[switch,modulo]{lineno}     

\newcommand{\xbj}{x_{\text{Bj}}}
\newcommand{\xf}{x_{\text{F}}}
\newcommand{\ubar}{\bar u}
\newcommand{\dbar}{\bar d}
\newcommand{\jpsi}{J/\psi}

\newcommand{\us}{$\mu$s\xspace}

\newif\ifpdf
\ifx\pdfoutput\undefined
  \pdffalse
\else
  \pdfoutput=1
  \pdftrue
\fi
\ifpdf
  \usepackage{graphicx}
  \usepackage{epstopdf}
\else
  \usepackage{graphicx}
\fi
\begin{document}

\renewcommand*{\thefootnote}{\fnsymbol{footnote}}
\begin{frontmatter}
  \title{The SeaQuest Spectrometer at Fermilab}
  
\author[Michigan,LANL]{C. A. Aidala} 
\author[ANL]{J. R. Arrington} 
\author[Michigan]{C. Ayuso} 
\author[ACU]{B. M. Bowen} 
\author[ACU]{M. L. Bowen} 
\author[ACU]{K. L. Bowling} 
\author[ACU]{A. W. Brown} 
\author[FNAL]{C. N. Brown} 
\author[ACU]{R. Byrd} 
\author[ACU]{R. E. Carlisle} 
\author[Sinica]{T. Chang} 
\author[Sinica]{W.-C. Chang} 
\author[Illinois,Sinica,Maryland,Michigan]{A. Chen} 
\author[NSRRC]{J.-Y. Chen} 
\author[FNAL]{D. C. Christian} 
\author[LANL,FNAL]{X. Chu\footnote{Now at School of Physics, Peking University, Beijing 100871, China}} 
\author[Illinois]{B. P. Dannowitz} 
\author[ACU]{M. Daugherity} 
\author[Illinois]{M. Diefenthaler\footnote{Now at Thomas Jefferson National Accelerator Facility, Newport News, Virginia USA\label{fn:JLab}}} 
\author[Illinois]{J. Dove} 
\author[ANL]{C. Durandet\footnote{Now at Department of Physics, Paradise Valley Community College, Phoenix, Arizona USA}} 
\author[MississippiState,Rutgers]{L. El Fassi} 
\author[Colorado]{E. Erdos} 
\author[ACU]{D. M. Fox} 
\author[ANL]{D. F. Geesaman} 
\author[Rutgers]{R. Gilman} 
\author[RIKEN]{Y. Goto} 
\author[LANL]{L. Guo\footnote{Now at Florida International University, Miami, Fl 33199 USA}} 
\author[Kaohsiung]{R. Guo} 
\author[ANL,ACU]{T. Hague} 
\author[ACU]{C. R. Hicks} 
\author[ANL]{R. J. Holt\footnote{Now at Kellogg Radiation Laboratory, California Institute of Technology, Pasadena, California 91125 USA}}  
\author[ACU]{D. Isenhower} 
\author[LANL]{X. Jiang} 
\author[Colorado]{J. M. Katich} 
\author[Illinois]{B. M. Kerns} 
\author[Colorado]{E. R. Kinney} 
\author[ACU]{N. D. Kitts} 
\author[LANL]{A. Klein} 
\author[LANL]{D. Kleinjan} 
\author[Illinois]{J. Kras} 
\author[Yamagata]{Y. Kudo} 
\author[Colorado]{P.-J. Lin} 
\author[Illinois] {D. Liu} 
\author[LANL]{K.  Liu} 
\author[LANL]{M. X. Liu} 
\author[Michigan]{W. Lorenzon} 
\author[Illinois] {N. C. R. Makins} 
\author[ACU]{J. D. Martinez} 
\author[Illinois]{R. E. McClellan\textsuperscript{\ref{fn:JLab}}} 
\author[Colorado]{B. McDonald} 
\author[LANL]{P. L. McGaughey} 
\author[ACU]{S. E. McNease} 
\author[ANL]{M. M. Medeiros} 
\author[ACU]{B. Miller} 
\author[ACU]{A. J. Miller} 
\author[TokyoTech]{S. Miyasaka} 
\author[Yamagata]{Y. Miyachi} 
\author[Michigan]{I. A. Mooney} 
\author[Michigan]{D. H. Morton} 
\author[ANL,Michigan]{B. Nadim} 
\author[TokyoTech]{K.  Nagai} 
\author[Maryland]{K. Nakahara\footnote{Now at Stanford Linear Accelerator Center, Menlo Park, California 94025 USA}} 
\author[TokyoTech]{K. Nakano} 
\author[Yamagata]{S. Nara} 
\author[TokyoTech]{S. Obata} 
\author[Illinois]{J. C. Peng} 
\author[Illinois] {S. Prasad} 
\author[LANL]{A. J. R. Puckett\footnote{Now at University of Connecticut, Storrs, Connecticut, 06269, USA}} 
\author[Michigan]{B. J. Ramson} 
\author[Michigan]{R. S. Raymond} 
\author[ANL]{P. E. Reimer} 
\author[Michigan,ANL]{J. G. Rubin} 
\author[ACU]{R. Salinas} 
\author[TokyoTech]{F. Sanftl} 
\author[KEK]{S. Sawada} 
\author[Michigan]{T. Sawada} 
\author[Michigan]{M. B. C. Scott} 
\author[ACU]{L. E. Selensky} 
\author[TokyoTech]{T.-A. Shibata} 
\author[Sinica,NCUTaiwan]{S. Shiu} 
\author[Sinica]{D. Su} 
\author[Rutgers]{A. S. Tadepalli} 
\author[Illinois]{M. Teo\footnote{Now at Stanford University, Stanford, California 94305 USA}} 
\author[ANL]{B. G. Tice} 
\author[ACU]{C. L. Towell} 
\author[ACU]{R. S. Towell} 
\author[LANL]{S. Uemura} 
\author[Sinica,FNAL,Kaohsiung]{S. G. Wang\footnote{Now at ChemMatCARS, The University of Chicago, Argonne, Illinois 60439 USA}} 
\author[ACU]{S. Watson} 
\author[ACU]{N. White} 
\author[LANL]{A. B. Wickes} 
\author[Michigan]{M. R. Wood} 
\author[FNAL]{J. Wu} 
\author[ACU]{Z. Xi} 
\author[ANL]{Z. Ye} 
\author[Illinois] {Y. Yin} 

\address[ACU]{Abilene Christian University, Abilene, Texas 79699 USA}
\address[Sinica]{Institute of Physics, Academia Sinica, Taipei,11529, Taiwan}
\address[ANL]{Physics Division, Argonne National Laboratory, Argonne, Illinois 60439, USA}
\address[Colorado]{University of Colorado, Boulder, Colorado 80309, USA}
\address[FNAL]{Fermi National Accelerator Laboratory, Batavia, Illinois 60510, USA}
\address[Illinois]{University of Illinois at Urbana-Champaign, Urbana, Illinois 61801, USA}
\address[Kaohsiung]{Department of Physics, National Kaohsiung Normal University, Kaohsiung 824, Taiwan}
\address[KEK]{KEK, High Energy Accelerator Research Organization, Tsukuba, Ibaraki 305-0801, Japan}
\address[LANL]{Los Alamos National Laboratory, Los Alamos, New Mexico 87545, USA}
\address[Maryland]{University of Maryland, College Park, Maryland 20742, USA}
\address[Michigan]{University of Michigan, Ann Arbor, Michigan 48109, USA}
\address[MississippiState]{Mississippi State University, Mississippi State, Mississippi 39762, USA}
\address[NCUTaiwan]{Department of Physics, National Central University, Jhongli District, Taoyuan City 32001,Taiwan}
\address[NSRRC]{National Synchrotron Radiation Research Center, Hsinchu, 30076, Taiwan}
\address[RIKEN]{RIKEN Nishina Center for Accelerator-Based Science, Wako, Saitama 351-0198, Japan}
\address[Rutgers]{Rutgers, The State University of New Jersey, Piscataway, New Jersey 08854, USA}
\address[TokyoTech]{Tokyo Institute of Technology, Tokyo, Japan}
\address[Yamagata]{Yamagata University, Yamagata, Japan}

  \begin{abstract}
The SeaQuest spectrometer at Fermilab was designed to detect oppositely-charged pairs of muons (dimuons) produced by interactions between a 120 GeV proton beam and liquid hydrogen, liquid deuterium and solid nuclear targets.  The primary physics program uses the Drell-Yan process to probe antiquark distributions in the target nucleon.  The spectrometer consists of a target system, two dipole magnets and four detector stations.  The upstream magnet is a closed-aperture solid iron magnet which also serves as the beam dump, while the second magnet is an open aperture magnet.   Each of the detector stations consists of scintillator hodoscopes and a high-resolution tracking device.  The FPGA-based trigger compares the hodoscope signals to a set of pre-programmed roads to determine if the event contains oppositely-signed, high-mass muon pairs.
\end{abstract}
  \begin{keyword}
    SeaQuest \sep 
    E906 \sep
    Drell-Yan \sep 
    spectrometer \sep 
    $J/\psi$ \sep 
    muon
  \end{keyword}
\end{frontmatter}

\renewcommand*{\thefootnote}{\arabic{footnote}}
\setcounter{footnote}{0}
\begin{multicols}{2}
\tableofcontents
\end{multicols}
\vspace*{0.25in}\hrule\vspace*{0.25in}

\sloppy
\begin{twocolumn}
\section{Introduction to SeaQuest\label{sec:intro}}

The proton is composed of an effervescing sea of quarks, antiquarks and gluons.  Many of the properties of the proton can be attributed to the flavors of quark excess (with respect to the antiquarks); however, the strong force and the sea of quark-antiquark pairs that it creates are primarily responsible for the proton's mass. The SeaQuest experiment was designed to explore the flavor dependence of the proton's sea quarks and modifications of the sea quark structure when the proton is contained within a nucleus.   

Sensitivity to the sea quark distributions is achieved through the Drell-Yan reaction that necessarily involves antiquarks.  To leading order in the strong coupling constant, $\alpha_s$, the Drell-Yan is a pure electromagnetic annihilation of a quark in one hadron with an antiquark in a different hadron forming a massive virtual photon that decays into a detectable  lepton-antilepton pair. This process was first observed by J.H. Christenson {\it et al.}~\cite{PhysRevLett.25.1523, PhysRevD.8.2016}.  The features of the cross section were explained by S.D. Drell and T.-M. Yan~\cite{PhysRevLett.25.316, PhysRevLett.25.902.2} in terms of the parton model as a hard scattering of point-like partons multiplied by a convolution of the parton distributions of the interacting hadrons:
\begin{eqnarray}
\lefteqn{\frac{d^2\sigma}{dx_1dx_2} = \frac{4\pi\alpha_e^2}{9sx_1x_2}\times}\hspace{0.05in} \\ \nonumber
& \displaystyle\sum_{q\in\{u,d,\dots\}}  e_q^2\left[\bar q_1(x_1)q_2(x_2) + q_1(x_1)\bar q_2(x_2)\right],\label{eq:dyxs}
\end{eqnarray}
where $x_{i}$ represents Bjorken-$x$, $\xbj$, of the interacting parton in hadron $i$ (generally, the beam parton is denoted as 1 and the target as 2); $q_i(x_i)$ is the parton distribution of quark of flavor $q$; $e_q$ is the charge of quark flavor $q$; $\sqrt{s}$ is the center-of-mass energy; $\alpha_e \approx 1/137$ is the fine structure constant; and the sum is over all quark flavors.  The $e_q^2$ weighting of the parton distributions implies that with a proton beam, the cross section is primarily sensitive to the $u$- and $\ubar$-quark distributions.  This expression is only leading order, and next-to-leading order (NLO) terms with the first power of $\alpha_s$, contribute up to half of the cross section.  In the SeaQuest spectrometer and many typical fixed-target experiments, the acceptance is biased toward large, positive Feynman-$x$, $\xf \approx x_1 - x_2$, and thus the beam parton is generally a large $\xbj$ valence parton.  For a proton beam, this implies that the interaction is between a valence beam {\em quark} and a lower-$\xbj$ target {\em antiquark}. 

The SeaQuest experiment was proposed to measure the distributions of sea quarks in the nucleon, specifically the ratio of anti-down to anti-up quarks in the proton and the modifications of these distributions in nuclei (the anti-quark EMC effect).  In addition, despite the absence of initial state hadron polarization, a convolution of Boer-Mulders Transverse Momentum Dependent distributions is accessible through the azimuthal distributions of the virtual photon's decay products. 

The spectrometer was designed to measure the $\mu^+\mu^-$ (dimuon) decay of the Drell-Yan virtual photon, produced using a 120 GeV proton beam extracted in a \unit[4]{s} long, slow-spill from the Fermilab Main Injector.     The spectrometer's basic concept is based on previous Fermilab Drell-Yan spectrometers that were used with 800 GeV extracted beams~\cite{PhysRevD.43.2815, PhysRevLett.64.2479, PhysRevLett.80.3715, Towell:2001nh}.   To account for the difference in boost, the spectrometer was shortened significantly.   Key features of the spectrometer include two large dipole magnets and four independent tracking/triggering stations.  A schematic drawing of the spectrometer is shown in Fig.~\ref{fig:spect}.  The experiment uses liquid hydrogen, liquid deuterium, carbon, iron and tungsten targets as well as an empty liquid target flask and a ``no target'' position for background subtraction.  Only one target intercepts the beam during any given slow-extraction spill.   The first dipole magnet (called FMag) is a closed-aperture, solid iron magnet.  The beam protons that do not interact in the targets are absorbed in the iron of the first magnet, which allows only muons to traverse the remaining spectrometer.  The downstream magnet (denoted KMag) is a large, open-aperture magnet that was previously used in the Fermilab KTeV experiment~\cite{PhysRevD.67.012005}.  Each of the tracking/triggering stations consists of a set of scintillator hodoscopes to provide fast signals for an FPGA-based trigger system and a drift chamber for fine-grained tracking.  

\begin{figure*}[tb]
  \centering
  \includegraphics[width=\textwidth,clip]{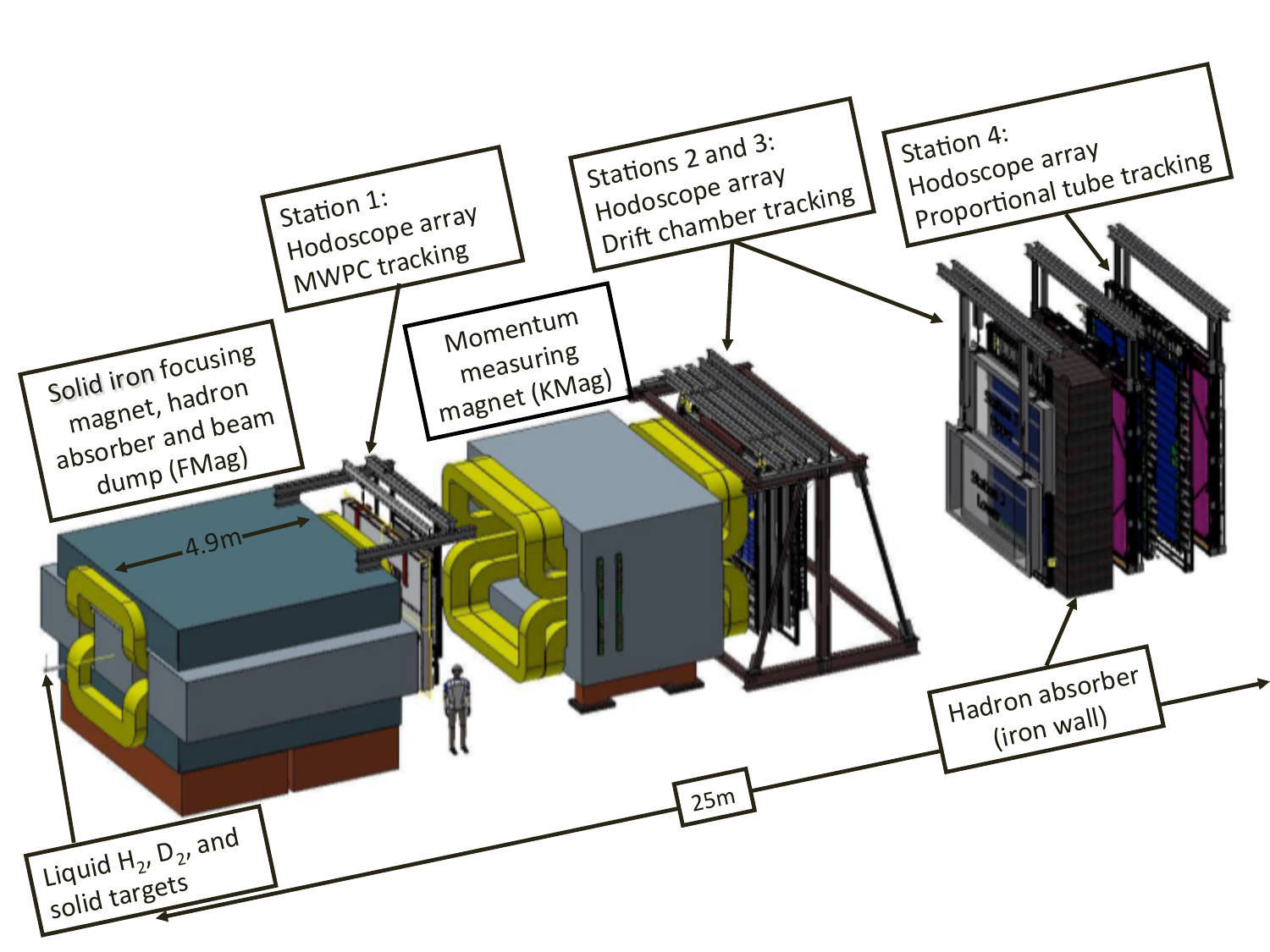}
  \caption{Schematic of the SeaQuest spectrometer.  The 120 GeV proton beam enters from the left, and the solid iron magnet also serves as an absorber for the beam that did not interact in a target.  The active elements of the spectrometer are shielded interactions of the primary proton beam by additional concrete and steel enclosing FMag and the target region that is not shown here.}
  \label{fig:spect}
\end{figure*} 

During the design of the spectrometer, the acceptance was extensively studied with Monte Carlo simulations.  The goal was to maximize the acceptance for events with large $x_2$ within the constraints of the existing equipment, most significantly KMag, the coils in FMag and the drift chambers at station 2.  The other primary optimization was the target to beam dump distance.  To distinguish between tracks from the target and the beam dump, this distance should be relatively long; however, this increases the single muon background from pions formed in the target that decay before reaching the dump.  The experiment uses a coordinate system where positive $z$ is along the proton beam direction, positive $y$ is vertically up and positive $x$ is to beam left to complete the right-handed system.

This article describes each element of the spectrometer and associated systems.  Over the course of the experiment, the spectrometer was upgraded several times.  The recorded data have been divided into ``data sets'' based on the specific configuration of the detector and trigger.  Table~\ref{tab:dataset} lists the dates when each data set was recorded and the major spectrometer changes between the data sets.

\newcommand{\mrw}[3]{\multirow{#1}{#2}{#3}}
\begin{table*}[tb]
\begin{center}
  \caption{The SeaQuest experiment's data sets and the dates when they were recorded.  Note that the major breaks generally correlate with the Fermilab accelerator maintenance periods. Live Prot. is the integral number of protons on target that were not vetoed by the BIM (Sec.~\ref{sec:BIM}) and while the DAQ was live.  Section~\ref{sec:tracking} explains the drift chamber configuration nomenclature.  \label{tab:dataset}}
  
   \begin{tabular}{cccll}
    \\ \hline\hline
                      &                                                       & Live 
                                                                                           &                                                       &\\
    Data          &                                                        & Prot.
                                                                                           &                                                       & \\
    Set            & Dates                                              & $\times 10^{17}$
                                                                                           & Drift Chamber Config.                   & Comments\\
    \hline\vspace{6pt}
   1                 & Mar.--Apr. 2012                              &        & DC1.1;DC2;DC3p-m.1                 &   Commissioning \\ 
\mrw{4}{*}{2} & \mrw{4}{*}{Nov. 2013--Sep. 2014}  & \mrw{4}{*}{2.0} & \mrw{4}{*}{DC1.1;DC2;DC3p-m.2} &  New station 3 (lower) drift \vspace*{-2pt} \\
                      &                                                       &         &              & \hspace{12pt}chamber\\
                      &                                                       &         &              &  New stations 1 and 2\vspace*{-2pt} \\ 
                      &                                                       &         &                                                           &   \hspace{12pt}photomultiplier bases\\
   3                 & Nov. 2014--Jul. 2015                     &  6.1   &  DC1.1;DC2;DC3p-m.2                     &  \\
\mrw{2}{*}{4} & \mrw{2}{*}{Nov. 2015--Feb. 2016} & \mrw{2}{*}{0.8}  & \mrw{2}{*}{DC1.2;DC2;DC3p-m.2} & New station 1 drift \vspace*{-2pt} \\
                      &                                                       &                          &                                                          & \hspace{12pt}chamber (DC1.2)\\ 
\mrw{2}{*}{5} & \mrw{2}{*}{Mar. 2016--Jul. 2016}   & \mrw{2}{*}{2.5}        &  DC1.1;DC1.2;                                  & Both DC1.1 and 1.2 installed\vspace*{-2pt} \\
                     &                                                        &         & DC2;DC3p-3m.2                               & \hspace{12pt}in station 1\\ 
\mrw{2}{*}{6} & \mrw{2}{*}{Nov. 2016--Jul. 2017}   & \mrw{2}{*}{ 2.3}      & DC1.1;DC1.2;                                   & \mrw{2}{*}{DAQ upgrade (See Sec.~\ref{sec:eventdaq}.)}\\
                      &                                                      &          & DC2;DC3p-3m.2 &  \\    \hline\hline
  \end{tabular}
  \end{center}
  \end{table*}
\section{Proton Beam Intensity Monitor\label{sec:beam}}

SeaQuest uses the 120 GeV proton beam from the Fermilab Main Injector.  The beam is extracted in a slow spill lasting just under four seconds.  Typically, the time between the beginning of spills is just over one minute.  Beam is extracted using a resonant process and the extracted beam retains the 53.1 MHz structure of the Main Injector RF, dividing the beam into ``RF buckets'' that are less than \unit[2]{ns} long and occur every \unit[18.8]{ns}.  

\subsection{Sensitivity to Instantaneous Intensity}
The SeaQuest beam intensity is not uniform in time throughout the slow spill.  There are beam buckets in the Main Injector that are intentionally left empty to allow the injection kickers to inject 8 GeV protons from the Fermilab Booster without disturbing the protons already in the Main Injector and to allow the abort kickers ramp to full field if needed.   Typically, 492 of the 588 RF buckets in the Main Injector contain protons during the SeaQuest slow spill cycle.  Unfortunately for SeaQuest, the number of protons in these 492 buckets varies greatly throughout a slow spill, as is shown in Fig.~\ref{fig:cherenkovPlot}.

\begin{figure*}
  \begin{center}
    \includegraphics[width=0.95\textwidth]{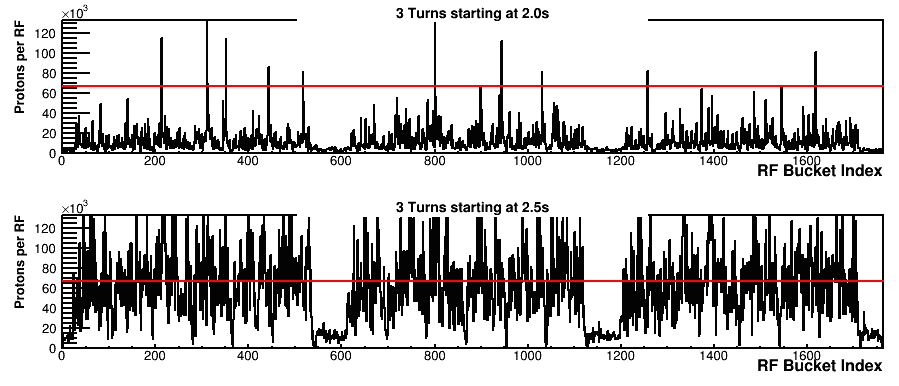}
    \caption{The beam intensity measured by the Beam DAQ Cerenkov counter every beam bucket.  Each strip shows the number of protons per beam bucket as a function of time for approximately \unit[33]{$\mu$s}.  The upper and lower plots begin \unit[2.0]{s} and   \unit[2.5]{s} after the beginning of the same spill.  The horizontal red line in each plot indicates the threshold above which the trigger is inhibited.  In the lower plot, a significant number of the ``RF buckets'' were above this threshold.  During this time and in the surrounding buckets the trigger was inhibited and no data was recorded. \label{fig:cherenkovPlot}}
  \end{center}
\end{figure*}

The SeaQuest trigger is synchronized with the Main Injector RF and is able to discriminate between muons from interactions in different RF buckets.   The SeaQuest trigger is designed to accept events containing a high mass pair of oppositely charged muons.  Typically, this implies muon pairs in which both tracks have high transverse momentum.  The trigger uses hits in one scintillation counter hodoscope located between FMag and KMag and three hodoscopes located downstream of KMag. For a detailed description of the  hodoscopes  and trigger,  see Secs.~\ref{sec:hodo} and \ref{sec:trigger}, respectively. However, the vast majority of SeaQuest triggers are the result of hits from a number of unrelated particle tracks that can mimic a high mass muon pair. The probability that this type of trigger will occur increases dramatically with  proton beam intensity. When this is combined with the non-uniformity of the slow spill extraction, the data acquisition system can be saturated with undesired triggers. The Beam Intensity Monitor was designed to solve this problem.

\subsection{Beam Intensity Monitor \label{sec:BIM}}
The SeaQuest Beam Intensity Monitor (BIM) senses when the beam intensity is above a (programmable) threshold and inhibits triggers for a window around the high-intensity RF bucket.  The duration of the inhibit window is programmable, and was typically set to $\pm 9$ RF buckets.  The inhibit threshold is generally set between 65,000 and 95,000 protons per RF bucket\footnote{At the proposed beam intensity of $5\times 10^{12}$ protons/(4 s) spill, the average number of protons in a full RF bucket is approximately 28,000.}.  The beam intensity is measured using a gas Cerenkov counter operated at atmospheric pressure, as shown in Fig.~\ref{fig:BIMCerenkov}.
\begin{figure}[tb]
	\begin{center}
		\includegraphics[width=\columnwidth]{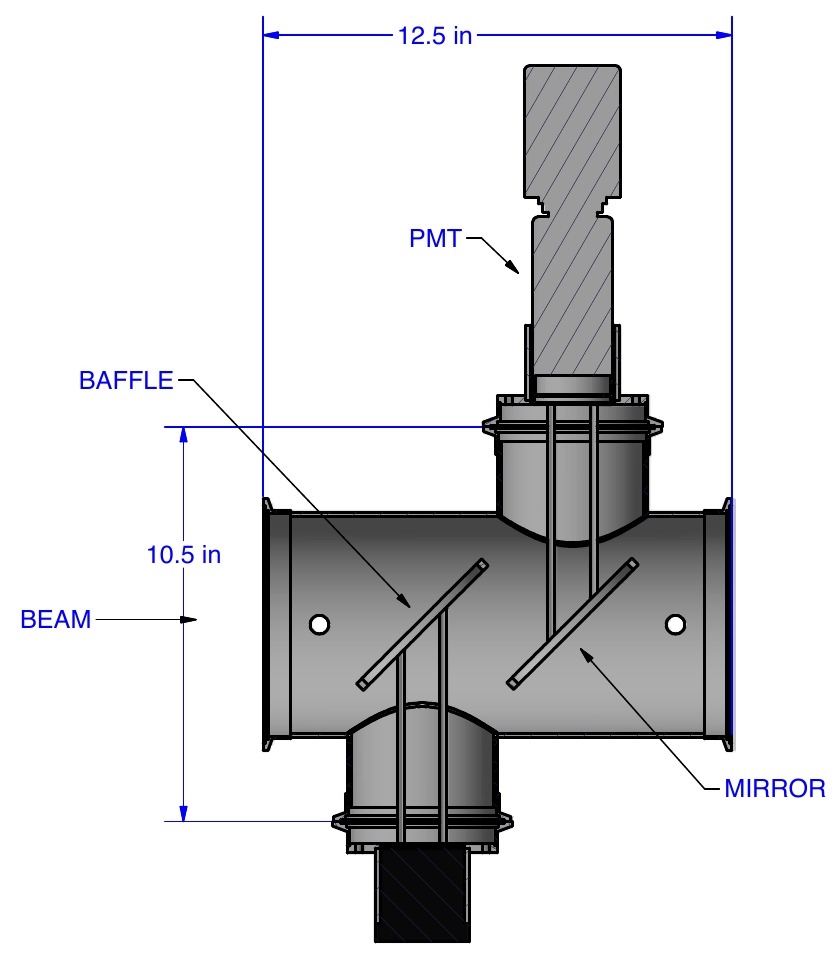}
		\caption{The Beam Intensity Monitor (BIM) Cerenkov counter.  Dimensions are given in inches.  The box at the bottom of the diagram is a cylinder of black plastic that covers the access port holding the mount for the black paper baffle.\label{fig:BIMCerenkov}}
	\end{center}
\end{figure}
The counter and digitization electronics were designed to have good time resolution, and a linear response over a large dynamic range.   A 45$^\circ$ aluminized Kapton\footnote{After exposure to approximately $3\times10^{17}$ protons, the mirror reflectivity is significantly reduced in the beam spot and the mirror is replaced.  The best mirror lifetime was found with a relatively thick vapor deposited layer of aluminum (1 $\mu$m).} mirror held on an elliptical G10 frame directs light to a single photomultiplier tube (PMT).  A baffle of black construction paper held parallel to the mirror ensures that the active path length in the radiator\footnote{A gas mixture of 80$\%$ Argon and 20$\%$ CO$_2$ is used as the Cerenkov radiator.  This gas mixture is used in the beam line instrumentation package located just upstream of the BIM and was chosen for convenience.} for protons is independent of beam position.  A two-inch diameter 8-stage photomultiplier tube (Electron Tubes 9215B) is positioned close to the mirror so that all Cerenkov light falls directly on the face of the phototube.  The phototube and ``fully transistorized'' voltage divider\footnote{The voltage divider (Electron Tubes Part Number TB1102C284AFN2) uses a circuit based on \cite{Kerns} to make the dynode voltages independent of phototube current.} (also provided by Electron Tubes) were chosen to maximize dynamic range.

The BIM photomultiplier tube signal is carried on an RG8 cable (approx. 50 ft. long) to a Fermilab-designed NIM module located outside of the high radiation area.  The signal is integrated and digitized using a custom integrated circuit designed at Fermilab for the CMS experiment at the CERN Large Hadron Collider. This chip is one of the ``QIE'' (Charge Integrator and Encoder) family of circuits used first by the KTeV experiment at Fermilab~\cite{QIE}. The chip is clocked with the Main Injector RF clock and provides an ADC conversion every 18.8 ns clock cycle.  The output is encoded using eight bits and a non-linear scale that provides approximately constant binning resolution (bin size divided by bin magnitude) over a dynamic range of $10^5$.  The QIE bin size contributes an RMS uncertainty in the measured beam intensity of approximately  $1\%$.
The light incident on the photomultiplier tube is attenuated using neutral density filters so that the QIE least count corresponds to about 30 protons per beam bucket.  The QIE full scale corresponds to more than $3\times 10^6$ protons per beam bucket. 

In addition to inhibiting triggers when the instantaneous intensity is above threshold, the BIM interface module provides the information required to count the number of protons incident on the SeaQuest targets while the experiment is ready and able to trigger.   The BIM interface module provides  (a) integrated beam for entire spill; (b) integrated beam while inhibit is asserted at trigger logic; (c) integrated beam during trigger dead time, excluding buckets inhibited while the event is being recorded; (d) a snapshot of beam intensity close in time to the trigger (ADC measurements for 16 buckets before and after the trigger and the triggered bucket); and (e) a complete record of the bucket-by-bucket intensity for the slow spill.  The timing of the inhibit signal and of all of the sums calculated by the BIM interface module are controlled using programmable registers.  The module is normally controlled using a 100~Mbps Ethernet interface.  One third of the complete spill record is recorded through the same Ethernet interface used to control the module.  Two additional 100~Mbps Ethernet interfaces are used to record the remainder of the complete spill information.  This readout occurs between spills.  The snapshot of beam intensity close in time to the trigger is also output on a twisted-pair ribbon cable and is recorded through the SeaQuest event data acquisition system.

The linear range of the phototube and voltage divider was established using an LED pulser.  The largest (linear) dynamic range was found with a bias voltage of about -900~V.  This agrees with vendor-provided information on the phototube performance.  The neutral density filters used to attenuate the Cerenkov light allow the tube to be biased at -870~V while providing signals of appropriate amplitude to match the QIE dynamic range.

The BIM measurement of beam intensity is normalized using a Secondary Emission Monitor (SEM) located upstream of the Cerenkov counter.  The SEM signal is integrated over each spill.  It is calibrated by measuring the activation of a thin foil placed in the beam.  The linear dynamic range of the BIM measurement was also verified using the SEM.
data set\section{Cryogenic and Solid Targets\label{sec:targets}}

The SeaQuest targets are centered 130 cm upstream of the first spectrometer magnet.  The general design and many parts of the target are inherited from the E866/NuSea experiment~\cite{PhysRevLett.80.3715, Towell:2001nh}.   As depicted in Fig. \ref{fig:target}, the target system consists of two liquid targets, three solid targets, and two positions for measuring background count rates--an empty flask and an empty solid-target holder.  The targets are mounted on a remotely positionable table which translates in the $x$-direction over a range of 91.4 cm. 
\begin{figure}[tb]
	\centering
	\includegraphics[width=\columnwidth]{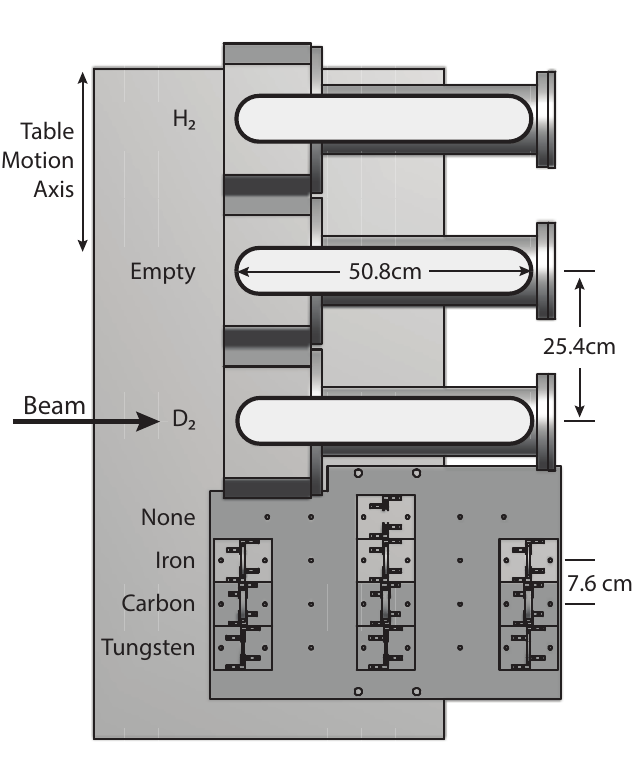}
	\caption{Top view of schematic layout of movable target table showing the seven target positions
        \label{fig:target} }
\end{figure}

The details of the target materials in the seven target positions are summarized in Tab. \ref{tab:target-materials}. The $^1H_2$ target used commercially available ``Ultra High Purity'' gas which was 99.999\% pure.  The deuterium came from two different sources.  The first is a supply of gas at Fermilab that was previously used in bubble chamber experiments.  This gas was known to have a small hydrogen contamination and was measured by mass spectroscopy to be $95.8\pm0.2\%$ $^2H$ with $^1H$ as the balance, primarily in HD molecules.  Later, SeaQuest switched to commercially available $^2H_2$ that had a purity of  99.90\% with $N_2$ as the balance.  The $N_2$ would have condensed in the pre-target cold trap.
\begin{table*}[t]
  \begin{center}
\caption{Characteristics of the seven SeaQuest target positions.  The ``Spills/Cycle'' should be regarded as a typical configuration.  It can vary in response to sample balancing needs and running configurations.  The non-zero interaction length of the empty flask is due to the 51$\mu$m-thick stainless steel end-caps of the flask and the 140 $\mu$m-thick titanium windows of the vacuum vessel that contains it. \label{tab:target-materials}}
    \begin{tabular}{c c c c c c c}
    \\ \hline\hline
               &               &                    &                  & Number of  & \\
               &               & Density       & Thickness & Interaction  & Spills/\\
 Position & Material & (g/cm$^3$) & (cm)           & Lengths  & Cycle \\ \hline
	1 &    $H_2$  & 0.071      & 50.8 & 0.069 &10 & \\
	2 & Empty Flask& --  & --  & 0.0016 &2 & \\
	3 &    $D_2$    &0.163    & 50.8 & 0.120&5 & \\
	4 & No Target    & --  & -- &  0 & 2 & \\
	5 &    Iron          &7.87  & 1.905 & 0.114 &1    & \\
	6 & Carbon       &1.80  & 3.322 & 0.209 &2  & \\
	7 & Tungsten    &19.30  & 0.953 & 0.096 & 1 & \\ [0.5ex] \hline\hline
    \end{tabular}		
  \end{center}
\end{table*}

Each of the solid targets is divided into three disks of 1/3 the total thickness listed in Table~\ref{tab:target-materials}.  These are spaced 25.4 cm apart along the beam axis to approximate the spatial distribution of the liquid targets, thereby minimizing target-dependent variation in spectrometer acceptance.  The one exception to this is that during the data set 2 period the iron disks were more closely spaced (17.1 cm).

\subsection{Target Control and Motion}
The control system for the cryogenic targets uses a Siemens APACS+ programmable logic controller (PLC).  This system contains several modules providing a large number of analog and digital input and output channels.  Nearly all of the sensors providing telemetry on table position and liquid target parameters are processed by this system and the majority of the signals controlling valves, feedback for heating systems, and power signals for pumps and refrigerators originate in this system.  The PLC is powered by an uninterruptible power supply and is capable of regulating the target systems and taking action under a large number of problem scenarios, even if disconnected from the target control computer and the rest of the network.  The target control computer communicates with the PLC via an ``M-BUS'' interface.  Programming and configuration of the PLC code is performed with Siemens 4-Mation software, and the real-time user interface to the PLC is built using the GE Fanuc iFix suite of software.  The graphical user interface is built in iFix Workspace.  This suite also includes remote historical data warehousing and plotting through iFix Historian and Proficy Portal software.

Motion of the target table is accomplished with a stepper motor driving a lead screw which moves the table on rails.  The stepper motor, motor driver, and motor controller are made by Anaheim Automation.  A single step of the motor translates the target table by 2.54~$\mu$m and target positions are confirmed by monitoring magnetic proximity switches mounted to the translating table and platform base.  The software step position is recalibrated to the edge of the central proximity sensor each time the table passes.  The motor controller is programmed using Anaheim Automation SMC60WIN software, running on the target control computer, and connected via USB.  In operation, the target controller requires only control signals sent to its input registers from the PLC to reposition the target.  Autoradiography of the titanium windows and solid target disks has shown the beam to be positioned within 5mm of target center and well within the target area for all positions.

\subsection{Cryogenic Target System}
The target flasks are 50.8 cm in length and 7.62 cm in diameter and each contains 2.2 l of liquid.   The flask walls are made of 76 $\mu$m-thick stainless steel with 51$\mu$m-thick stainless steel end-caps.  Each flask is tested to a gauge pressure exceeding 110 kPa and is leak-checked to better than $10^{-9}$~scc/s. The hydrogen and deuterium targets are liquefied from bottled gas by a pair of closed circuit He refrigeration systems.  Each refrigerator is a Cryomech water-cooled CP950 compressor and AL230 cold head (Gifford-McMahon cycle) capable of approximately 25 W of cooling power at 20 K.  The hydrogen and deuterium targets take approximately 18 and 12 hours to fill, respectively.  Temperature sensitive resistors are used to monitor the level of the liquid during filling and data-taking.

In order to maintain liquid in the flask and control its density, the targets are operated along the vapor-liquid saturation curve.  The pressure of vapor in the lines at the top of the flask is measured and that pressure is used to regulate power delivered to a group of three parallel, \unit[500]{$\Omega$} heater resistors.  A Watlow silicon controlled rectifier, controlled by the PLC, regulates the fraction of time 75 V is applied to the resistors, producing as much as 31.1 W of integrated heater power.  A desired gas pressure (typically just above atmospheric pressure)  is selected and the PLC regulates the heater current appropriately via a PID (proportional-integral-derivative) loop.  The liquid density is computed using the intercept of the known saturation curve with the measured pressure.  Variations in pressure and temperature measurements are used to estimate the uncertainty in the density.

An insulation vacuum that surrounds each target flask greatly reduces the heat-load seen by the cryogenic refrigerator.  A schematic layout of the flask and high vacuum plumbing is given in Fig. \ref{fig:target2}.  During normal operation, insulation vacuum is maintained by a diffusion pump backed by a mechanical fore pump at a level better than a millipascal ($10^{-5}$  Torr).  A mechanical rough pump is used for purging the target flask prior to filling and as a fall-back mechanism should conditions preclude the use of the diffusion pumping system (e.g. badly spoiled vacuum or failure of the diffusion fore pumps). Flask pressure is monitored by a pair of redundant Setra pressure transducers on the supply and vent lines.  Cold-head temperatures are monitored by triple-redundant Cernox temperature sensors.  Fore and rough vacuum are measured by thermocouple gauges or convection vacuum gauges. 

\begin{figure}[tb]
	\centering
	\includegraphics[width=\columnwidth]{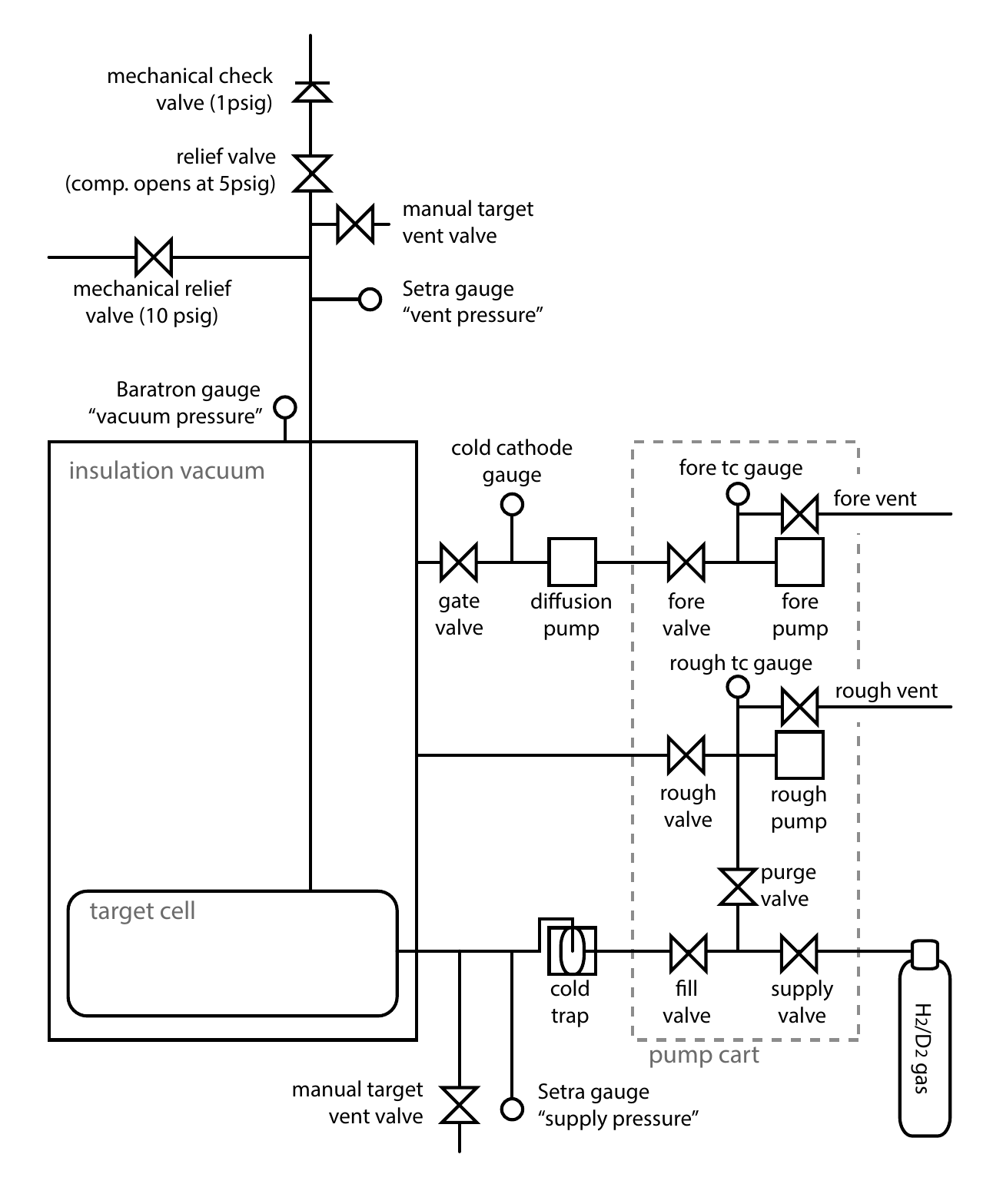} 
	\caption{Schematic layout depicting flask and vacuum plumbing of one of the two cryogenic targets. \label{fig:target2} }
\end{figure}

Figure \ref{fig:target3} shows the parameters of the liquid $D_2$ target as it is cooled down. The red curve shows the temperature of the condenser as it is cooled down from room temperature to 22 K. The blue curve indicates the resistance of the level sensor inside the target flask. The increase in the resistance is an indication of the formation of liquid inside the flask.

\begin{figure}[tb]
	\centering
	\includegraphics[width=\columnwidth]{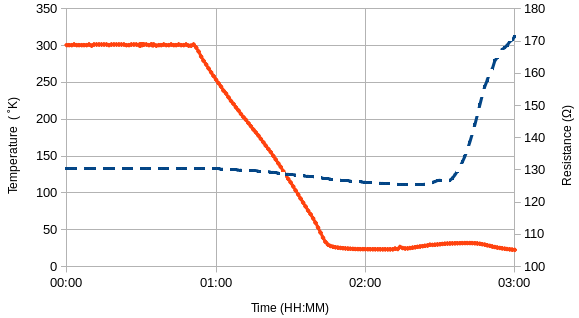} 
	\caption{The temperature of the condenser (red, solid curve, left scale) as the $D_2$ target being cooled down.  The blue, dashed curve (right scale) gives the resistance of the level sensor inside the target flask. \label{fig:target3}}
\end{figure}

 \section{Spectrometer Magnets \label{sec:magnets}}
 
 Central to the spectrometer are two large dipole magnets.  The upstream magnet (FMag) is a solid iron magnet assembled from 43.2 cm $\times$ 160 cm $\times$ 503 cm iron slabs, as shown in Fig.~\ref{fig:FMag}. The iron was recovered from the dismantled Columbia University Nevis Laboratory Cyclotron in 1980~\cite{NevisCyclotron}.  This iron is extremely pure, allowing a 2000 A excitation current (at 25 V, using 50 kW of power) to generate a magnetic field of 1.8 Tesla (yielding an average 3.07 GeV/c total magnetic deflection). FMag uses one of the three sets of ``bedstead'' coils recovered from the dismantled E866 SM3 magnet~\cite{PhysRevD.43.2815, PhysRevLett.64.2479, PhysRevLett.80.3715, Towell:2001nh}.  The coils are made from 5 cm square extruded aluminum.  The current exciting FMag is monitored by the Fermilab accelerator control system and is broadcast to the SeaQuest slow data acquisition system every acceleration cycle.  The magnet current is also input to the safety system to prevent beam from hitting the E906 spectrometer when FMag is not energized to a minimum level. The magnetic field distribution in FMag was modeled using a magnetostatic modeling program (EM Studio by Computer Simulation Technology).   The excitation was checked by wrapping a 1-turn coil around the central region of the magnet and integrating the charge induced as the magnet was energized.  The final calibration of the magnetic field is determined by examining the reconstructed mass of the $\jpsi$(3097) resonance. FMag acts as a spectrometer magnet and also as the beam dump for the 120 GeV beam delivered to the SeaQuest spectrometer.  There is a 5 cm diameter by 25 cm deep hole drilled into the upstream end of FMag along the beam axis.  The 120 GeV protons in the beam that do not interact in the SeaQuest targets interact in the central iron slab.  The hole moves the initial interaction points of the proton beam further away from the targets.  Most of the instantaneous 2.0 kW beam power is dissipated in this slab and is eventually conducted to the coils and the external surfaces of FMag to be radiated away.

\begin{figure}[tb]
  \centering
  \includegraphics[width=\columnwidth]{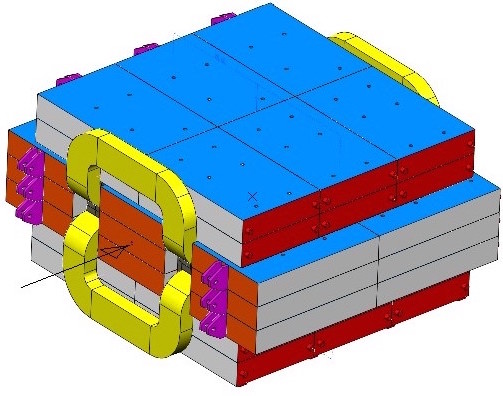}
	\caption{Perspective drawing of FMag showing the arrangement of the iron slabs. \label{fig:FMag}}
\end{figure}

The downstream magnet (KMag) is a 300 cm long iron rectangular magnet with a 289 cm wide by 203 cm high central air gap.  It was originally constructed by the E799/KTeV collaboration~\cite{PhysRevD.67.012005} at Fermilab, using donated steel from the University of Maryland Cyclotron for use in the E690 experiment.  It is excited to a central field of 0.4 Tesla (\unit[0.39]{GeV/c} magnetic deflection) by \unit[1600]{A} at \unit[270]{V} (\unit[430]{kW}).  The spatial distribution of the magnetic field in KMag was measured by the KTeV collaboration.  SeaQuest checked the central field calibration with a Hall probe.  The final value for the magnetic field again derives from the measurement of the exact mass of the $\jpsi$(3097) resonance.   The field of both magnets is oriented vertically so that the bend plane is horizontal.  In normal running conditions, both FMag and KMag bend muons in the same direction.  
\section{Hodoscopes\label{sec:hodo}}
Four plastic scintillator hodoscope stations are used for the primary trigger for the spectrometer.  Stations 1 and 2 use recycled scintillator bars from the HERMES experiment~\cite{Ackerstaff1998230}.  Stations 3 and 4 use new  Eljen EJ-200 scintillator material.   The x-planes are arranged vertically to measure the x-position (bend plane). The y-planes are arranged horizontally to measure the y-position (non-bend plane).

Stations 1 and 2 each have single x-y planes with 1 inch PMTs. Station 3 has a single x plane, and station 4 has two y planes and one x plane, all with 2 inch PMTs. The station 4 hodoscopes have PMTs on both ends of the scintillator bars. The scintillator bars in each plane are split to form a top/bottom or left/right pair. The bars are arranged with a slight (\unit[2-3]{mm}) overlap.  While improving efficiency, the overlap decreases the trigger's ability to reduce background by effectively making the trigger roads wider. Table~\ref{tab:hodo:planes} gives the number of scintillators, their physical sizes, and total apertures for each plane.  
\begin{table*}[t]
  \begin{center}
    \caption{Number and sizes of scintillators at each hodoscope plane.  Location refers to the distance along the beam axis from the front face of FMag. The designation (L) and (R) refer to beam left or right. The average efficiencies for the x-measuring trigger planes are also tabulated.  \label{tab:hodo:planes} 
    }
      \begin{tabular}{lcrrrrr@{\hskip3pt}l@{\hskip3pt}r}
       \\ \hline \hline
       \multicolumn{1}{c}{ }    & \multicolumn{1}{c}{ }    & \multicolumn{1}{c}{}   & \multicolumn{1}{c}{ }   & \multicolumn{1}{c}{ }   & \multicolumn{1}{c}{Array}  & \multicolumn{2}{c}{ }   \\ 
       \multicolumn{1}{c}{ }    & \multicolumn{1}{c}{ }    & \multicolumn{1}{c}{Length}   & \multicolumn{1}{c}{Width}   & \multicolumn{1}{c}{Thickness} & \multicolumn{1}{c}{Width}  & \multicolumn{2}{c}{Location}   & \multicolumn{1}{c}{Ave.}  \\ 
       \multicolumn{1}{c}{Plane}    & \multicolumn{1}{c}{Number}    & \multicolumn{1}{c}{(cm)}   & \multicolumn{1}{c}{(cm)}   & \multicolumn{1}{c}{(cm)}   & \multicolumn{1}{c}{(cm)} & \multicolumn{2}{c}{(cm)}  & \multicolumn{1}{c}{Eff.}     \\ \hline	
                1Y     & 20 $\times$ 2     & 78.7    &  7.32        & 0.64 &  140    &   663  \\
         1X      & 23 $\times$ 2    & 69.9    &  7.32        & 0.64 &161   &    653  && 0.978\\
         2Y     & 19 $\times$ 2     & 132.0    &   13.0        & 0.64 & 241   &  1403 \\
         2X     & 16 $\times$ 2     & 152.0     &   13.0        & 0.64 & 203   &  1421 && 0.989 \\
         3X     & 16 $\times$ 2     & 167.6     &   14.3        & 1.3 & 224   &  1958 && 0.959 \\  
\multirow{2}{*}{4Y1 }  & \multirow{2}{*}{16 $\times$ 2}  & \multirow{2}{*}{152.4}  &  \multirow{2}{*}{23.16}       &  \multirow{2}{*}{1.3}       & \multirow{2}{*}{366}&   2130 & (L)    \\
          &  &  & & & & 2146 & (R) \\
\multirow{2}{*}{4Y2 }  & \multirow{2}{*}{16 $\times$ 2} & \multirow{2}{*}{152.4}  &  \multirow{2}{*}{23.16}      &  \multirow{2}{*}{1.3}      & \multirow{2}{*}{366}      &  2200 & (L)    \\
          &  &  & & & & 2217 & (R)   \\ 
         4X     & 16 $\times$ 2     & 182.9  &   19.33    &  1.3 & 305      &  2240 && 0.979\\     \hline \hline
    \end{tabular}
  \end{center}
\end{table*}

Stations 1 and 2 must operate at very high rates and in the magnetic fringe fields from the large open-aperture magnet (KMag).  During the 2012 commissioning run, the stations 1 and  2 hodoscope occupancies were found to depend nonlinearly on the beam intensity, indicating that in the high rate sections of the hodoscopes, the phototube voltage dividers were sagging.  Before data taking resumed, a replacement voltage divider was designed and installed.  In the new design the voltage applied to dynodes $7-10$ of the phototube is stabilized using high voltage MOSFETs~\cite{Kerns}.  A unique feature of this design is that the current drawn by the base is constant, regardless of the rate seen by the phototube. The new voltage divider was designed to fit into the original metal package and reuses the original phototube sockets and external connectors.  The design is shown in Fig.~\ref{fig:hodo:st12base}.  Since the installation of the new voltage divider, no rate dependence has been observed in the stations 1 or 2 hodoscope occupancy plots.  

\begin{figure*}[tb]
  \begin{center}
     \includegraphics[width=0.65\textwidth]{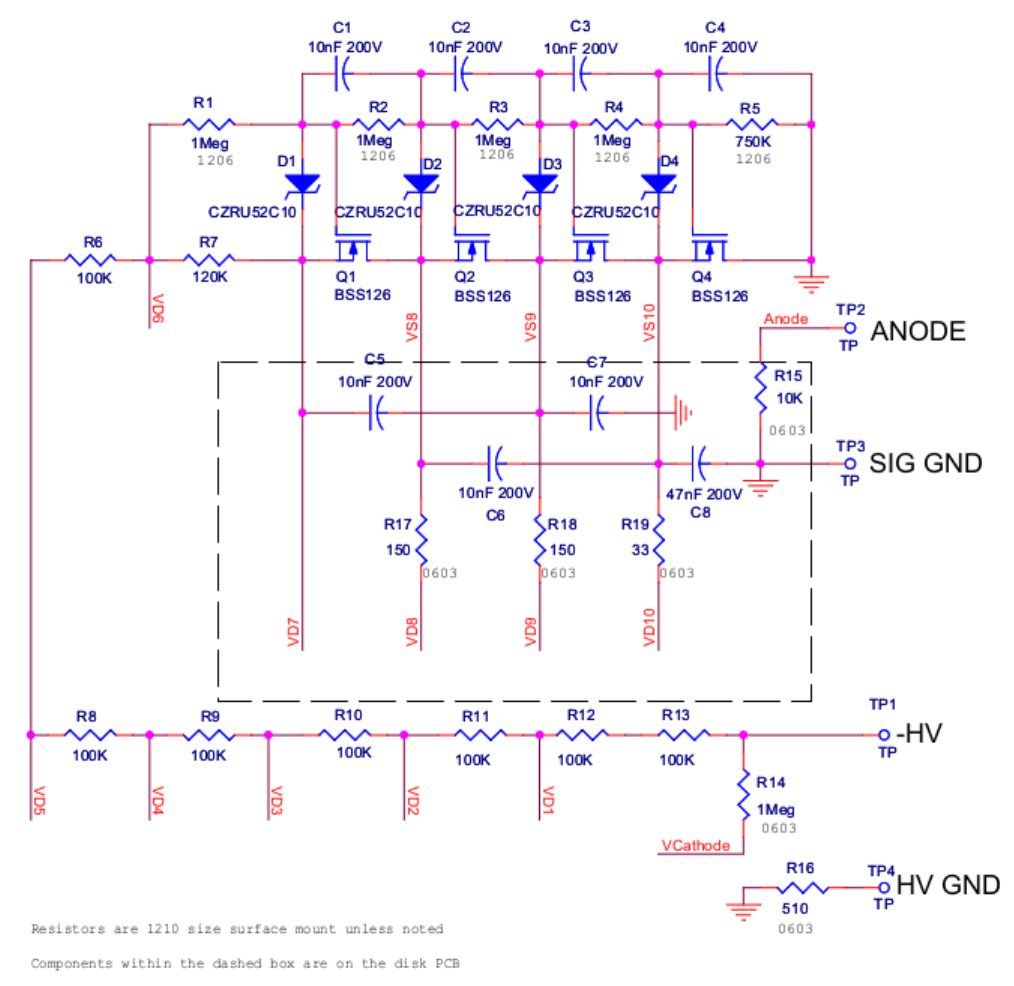}
    \caption{Schematic diagram of the MOSFET-stabilized photomultiplier base. \label{fig:hodo:st12base}}
  \end{center}
\end{figure*}

Due to the physical length of the scintillator bars, in particular in stations 3 and 4, the output pulses from the photomultipliers were 20-25 ns long. To correct for this each PMT base has a ``clip line'' attached at its output that has a 2.5ns long twisted pair cable with a 24 $\Omega$ resistor to provide an inverted reflected signal that arrives at the peak of the scintillator pulse to reduce the pulse width to approximately 10-15 ns full width.  The effect of the ``clip line'' can be seen in the oscilloscope traces in Fig.~\ref{fig:clipline}. All PMTs are powered by LeCroy 1440 High Voltage systems, which are controlled by software, which stores the voltages in a database, so that the setting for each run can be retrieved.  The hodoscope signal cables are $94\pm 2$ ns long. Each PMT signal is sent to a discriminator and then to a TDC.    Gross adjustments of the signal timing were made at a patch panel with short delay cables and fine adjustments were handled by the trigger FPGA modules (as discussed in Sec.~\ref{sec:trigger}).

To obtain the best trigger efficiency, the gains of the PMTs were carefully adjusted on a channel-by-channel basis.  Since the pulse height of the PMT signal was not recorded, this was accomplished by  observing the count rates over the range of allowable voltages for each PMT.  The operating voltage was set to be in the plateau of the voltage vs.\@ count rate curve.   This is done for each PMT as part of the regular maintenance of the system. The counter efficiency was checked using tracks reconstructed without requiring the given counter to have been hit either in the trigger or the reconstruction.  The plane averaged efficiencies for the hodoscopes used in forming the event trigger are listed in Tab.~\ref{tab:hodo:planes}.

\begin{figure}[tb]
  \begin{center}
      \includegraphics[width=\columnwidth]{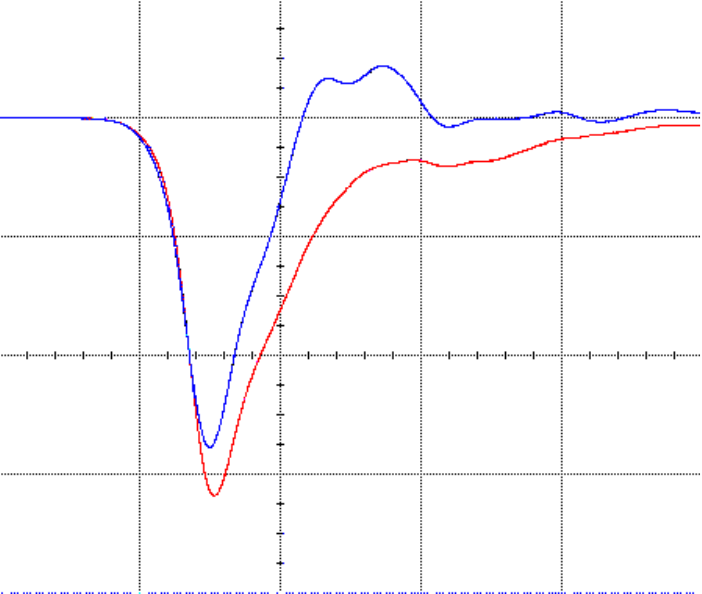}
    \caption{Oscilloscope traces recorded using a radioactive source on the station 2 hodoscopes without (red) and with (blue) the ``clip line".  Because of the reflection of the clip line, the amplitude of the blue curve is lower.  The vertical scale is 50 mV/div and the horizontal is 10 ns/div.  Both traces were averaged by the oscilloscope over 256 triggers. \label{fig:clipline}}
  \end{center}
\end{figure}

For maintenance, the hodoscope arrays in stations 3 and 4 can be rolled out of the beam line.  The frames holding the hodoscopes are suspended from above with a rail and ``jib traveler'' trolley system made by Ronstan\footnote{See \url{http://www.ronstan.com.au/marine5}.} for use in sailboats.  Each trolley can support 650 kg.  

data set\section{Tracking Chambers\label{sec:tracking}}

Drift chambers at stations 1, 2, and 3 measured the positions of muons at the stations  location. Each station contained two wire planes measuring the x-position as well as two planes each measuring left and right stereo angles at $\pm14^\circ$ (denoted u and v) for a total of six wire planes. The actual configuration of the chambers changed with time as will be discussed below. The resolution of dimuon invariant mass is dominated by multiple scattering and energy loss in FMag.  To keep the position resolution contribution to the overall resolution less than 10\% of the total mass resolution, the position resolution of an individual plane is required to be 400 $\mu$m, corresponding to a momentum resolution of $\Delta p / p$ (\%) $= 0.03 \cdot p$ (GeV/c).  To achieve a track reconstruction efficiency of at least 90\%, while allowing only one inefficient plane at every station, the single plane efficiency needed to be greater than 95\%.   The chambers must operate at high rates because of the large background particle flux.  This is most important at the most upstream chamber, station 1.  The hit rate is maximum at the edge of the chamber acceptance. As extracted from experimental data taken with an unbiased trigger, the average hit rate reaches a maximum of 3.0, 1.6 and 0.7 MHz/wire at stations 1, 2, and 3, respectively, at a beam intensity of $5\times10^{12}$ protons/(4~s).  It was necessary to design the chambers so that the probability of double hits per wire per event would be small.

\subsection{Chamber Configuration}
The basic structure is common to all the chambers. Each drift chamber consists of six planes of sense wires. Wires are aligned in the vertical direction in two planes called x and x$^\prime$. Wires are tilted by $+14^\circ$ in two planes called u and u$^\prime$, and by $-14^\circ$ in two planes called v and v$^\prime$. The wires in the primed planes are offset by half the drift cell size, which help to resolve the left-right ambiguity of drift direction. Every wire plane is oriented perpendicular to the $z$ axis, and each drift cell is rectangularly shaped. The drift chambers in the experiment are named ``DC'' followed by the station number, for example, at station 2  the chambers are called DC2.  At station 1, a smaller chamber DC1.1 was used for data sets 1-3. This was replaced by a new larger chamber DC1.2 with better expected high rate capability in data sets 4-6.  However a choice was made to reinstall DC1.1 upstream of its previous position for data sets 5-6, so for these periods, there were two drift chambers at station 1.   At station 3, separate drift chambers cover the top and bottom halves, and are called ``DC3p'' and ``DC3m'' where ``p'' and ``m'' stand for ``plus'' and ``minus.'' Table~\ref{table:cham:param} summarizes the parameters of the drift chambers. Over the course of the experiment DC3m was also upgraded, as indicated in Tab.~\ref{tab:dataset}.  With the exception of DC1.2, all chambers used a gas mixture of Ar:CH$_4$:CF$_4$ (88:8:4).  The gas mixture for DC1.2 is Ar:CF$_4$:C$_4$H$_{10}$:C$_3$H$_8$O$_2$ (81:5:12:2). With this gas mixture, the electron drift velocity of DC1.2 is larger than 50 $\mu$m/ns for essentially the entire cell.

\begin{table*}[tb]
  \begin{center}
  \caption{Parameters of all chambers.  Those of primed planes are almost the same as of unprimed planes. For the x measuring planes, the position is the distance between the chamber and the upstream face of FMag, while for u and v it denotes the offset relative to the x plane.\label{table:cham:param} }
  \begin{tabular}{cccccc}
    \\ \hline\hline
                  &           & No.           & Cell  & Width                  &  \\
                  &           & of & width & $\times$ height  & Position     \\ 
  Chamber & View & wires & (cm)  & (cm)                    &  (cm) \\ 
    \hline
    DC1.1    & x     & 160      & 0.64  & 102 $\times$ 122 &  616  \\
             & u, v  & 201      & 0.64  & 101 $\times$ 122 & $\pm$20 \\
    DC1.2    & x     & 320      & 0.50  & 153 $\times$ 137 &  691    \\
             & u, v  & 384      & 0.50  & 153 $\times$ 137 & $\pm$1.2 \\
    DC2      & x     & 112      & 2.1   & 233 $\times$ 264 & 1347    \\
             & u, v  & 128      & 2.0   & 233 $\times$ 264 & $\pm$25 \\
    DC3p     & x     & 116      & 2.0   & 232 $\times$ 166 & 1931    \\
             & u, v  & 134      & 2.0   & 268 $\times$ 166 & $\pm$6  \\
    DC3m.1   & x     & 176      & 1.0   & 179 $\times$ 168 & 1879    \\
             & u, v  & 208      & 1.0   & 171 $\times$ 163 & $\pm$19 \\
    DC3m.2   & x     & 116      & 2.0   & 232 $\times$ 166 & 1895    \\
             & u, v  & 134      & 2.0   & 268 $\times$ 166 & $\pm$6  \\
    \hline\hline
  \end{tabular}
  \end{center}
\end{table*}

The DC1.1, DC2 and DC3m.1 chambers have been used in previous Fermilab experiments,  E605 (DC2 and DC3m.1)~\cite{PhysRevD.43.2815} and E866/NuSea (DC1.1)~\cite{PhysRevLett.80.3715, Towell:2001nh}. Since these chambers had not been in use for over a decade, they required significant work to bring them to a working condition.  This process included restringing a large number (approximately 30\% of the sense wires) of broken or loose wires with wires of appropriate tension and developing new readout electronics.

The DC3p and DC3m.2 chambers were designed and constructed for this experiment in order to cover the large acceptance required at station 3. The initial commissioning data and data set 1 were completed with DC3m.1 while DC3m.2 was being designed and constructed.  As DC3m.2 is 25 cm wider than DC3m.1 at each side, it provided a 20\% increase in number of accepted events at $x_2 \approx 0.3$ and 10\% at $x_2 \approx 0.4$. DC3m.2 also provided better operational stability; a lower number of dead or noisy wires and lower leak current.  The DC1.2 chamber was also designed and constructed for this experiment. It is wider than DC1.1 by 25 cm at each side, also providing an increase in statistical sensitivity at large $x_2$. In addition, DC1.2 has a smaller cell width for better hit-rate tolerance.

\subsection{Electronics}

The SeaQuest wire chambers use a custom amplifier-discriminator integrated circuit called ASDQ designed at the University of Pennsylvania for CDF \cite{ASDQ}.   A new 8-channel ``ASDQ card'' and 64-channel ``Level Shifter Board'' (LSB) were developed at Fermilab for SeaQuest. The LSBs convert the differential hit signals output by the ASDQs to standard low-voltage differential signaling (LVDS).  They also provide bias, threshold, and control voltages for the ASDQ cards.

The ASDQ cards are mounted directly on each wire chamber.  On DC2 and DC3, copper grounding card guides\footnote{Unitrack ``Ground-R-Guide,'' see \url{http://www.unitrack.com/metal\_card\_guide-ground-r-guide.html}.} connect to the chamber frames and provide both mechanical support and the reference voltage for the ASDQ inputs.  Except for DC1.1, all cathode wires are biased with negative high voltage, and the anode wire signals are DC coupled to the ASDQ inputs.  DC1.1 uses positive HV applied to the anode wires.  The signals from these chambers are AC coupled to the ASDQ inputs through high voltage blocking capacitors that are integral to the chambers.

A single twisted pair ribbon cable connects each ASDQ card to an LSB. Signals from 8 ASDQ cards may be input to a single LSB.   The LVDS outputs from an LSB are carried on four 17-pair ribbon cables to TDC modules.  Typically, one 64-channel LSB is connected to one corresponding 64-channel TDC. Up to 14 LSBs are housed in a 9U crate that distributes \unit[24]{V} power to the LSBs on a narrow backplane.  The \unit[24]{V} is provided by a rack-mounted linear power supply connected to the LSB crate backplane.  Multiple LSB crates can be powered by a single supply.  One LSB per station is controlled using an Ethernet interface.  Control of the other LSBs at that station use a serial protocol implemented with short RJ11 cables linking one LSB to the next.  A test pulse can also be injected to selected groups of channels.  The timing of the test pulse can be determined by a TTL pulse distributed to the LSBs using a daisy-chain of LEMO cables, or independently by each ASDQ card being tested.

\subsection{Drift Chamber Performance}

The performance of each drift chamber was measured {\it in situ}.  Muon tracks were reconstructed with chamber planes of stations 1, 2, and 3. The efficiency for each plane was determined by measuring the probability of a hit near where the muon track crossed the plane.  The distance between hit and muon track (residual) was used to extract the position resolution. Measurement results for the chambers used in the running period April 2014 - June 2015 are summarized in Tab.~\ref{tab:cham:performance}. The single-plane efficiency is better than the requirement ($>$95\%) for all planes.  The position measurement resolution of all the planes is better than the requirement ($<$400 $\mu$m). The residuals from track fits are shown in Fig.~\ref{fig:res_chambers}.

\begin{table}\centering
  \caption{Performance of the drift chambers in the SeaQuest experiment between April 2014 and June 2015. The position resolution and detection efficiency are the average of the resolutions and efficiencies for all six planes in the specified chamber.  The resolutions are the average value for each chamber of the RMS of the difference between the measured position in a plane and the position calculated at the z coordinate of that plane using a fit with the plane excluded from the fit.
\label{tab:cham:performance}}
  \begin{tabular}{lrrr}
    \\ \hline \hline
     & Max.   & Pos. & Detection \\
            & drift  & res.     & eff. (\%) \\
 Chamber    & (ns)       & ($\mu$m)       & (min.-max.) \\ 
    \hline
    DC1.1   &     100    &   225          & 99-100 \\
    DC2     &     260    &   325          & 96-99 \\
    DC3p    &     220    &   240          & 95-98 \\
    DC3m.2  &     210    &   246          & 97-98 \\
    \hline\hline
  \end{tabular}
\end{table}

\begin{figure*}
  \begin{center}
    \includegraphics{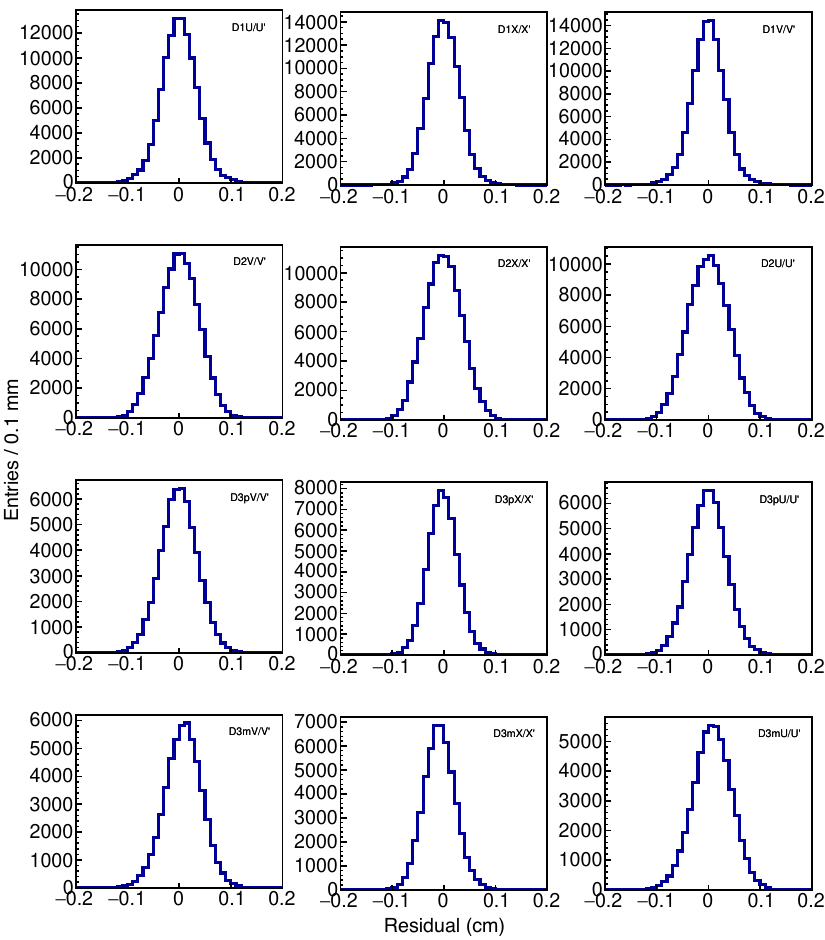}
    \caption{The residual distributions from each of the drift chamber planes. \label{fig:res_chambers}}
  \end{center}
\end{figure*}
The performance described above was measured with low beam intensity, specifically with fewer than $10^4$ protons per RF bucket, or approximately one quarter that of the normal intensity.  At normal intensity, the efficiency may be lower for several reasons.  For example, the probability that a signal hit is associated with a background hit is estimated to be 30\%, 40\% and 10\%\footnote{Probability = background rate $\times$ time window.  3.0 MHz/wire $\times$ 100 ns in the case of DC1.1} at the edges (where the hit rate is highest) of stations 1, 2 and 3, respectively.  Signal hits are sometimes lost due to associated background hits at the data-acquisition stage or the track reconstruction stage. This effect may cause a few percent of inefficiency at the edges of the stations. The rate tolerance of the chambers and analysis techniques to overcome possible limitations are being examined. 
\section{Muon Identification\label{sec:proptube}}

Muon identification at SeaQuest is accomplished with station 4, which is located downstream of a 1 m thick iron wall.  Like the other stations, this station contains both triggering hodoscopes (which were described in Sec.~\ref{sec:hodo}) and tracking detectors.  The station 4 tracking  detectors consist of 4 layers of proportional tube planes. Each plane is made of 9 proportional tube modules, with each module assembled from 16 proportional tubes, each \unit[12]{ft} (\unit[3.66]{m}) \footnote{English units are indicate that the fabrication was specified in English units.  (Metric units are provided for convenience as well.)} long with a \unit[2]{in} (\unit[5.08]{cm}) diameter, staggered to form two sub-layers. The proportional tubes are oriented along the horizontal (vertical) direction to provide precision measurements in the y(x)-coordinate in the first and fourth (second and third) planes, as shown in Fig.~\ref{fig:proptube:view}. The wall thickness of each tube is \unit[1/16]{in} (\unit[0.16]{cm}). The central anode wire is a gold-plated \unit[20]{$\mu$m} diameter tungsten wire. The proportional tubes used the same gas mixture as the drift chambers. The proportional tubes modules were originally developed for a Homeland Security project at Los Alamos National Laboratory that used cosmic ray muon radiography imaging~\cite{HomeLandSecurity}. 
\begin{figure}[tb]
  \centering
    \includegraphics[width=\columnwidth]{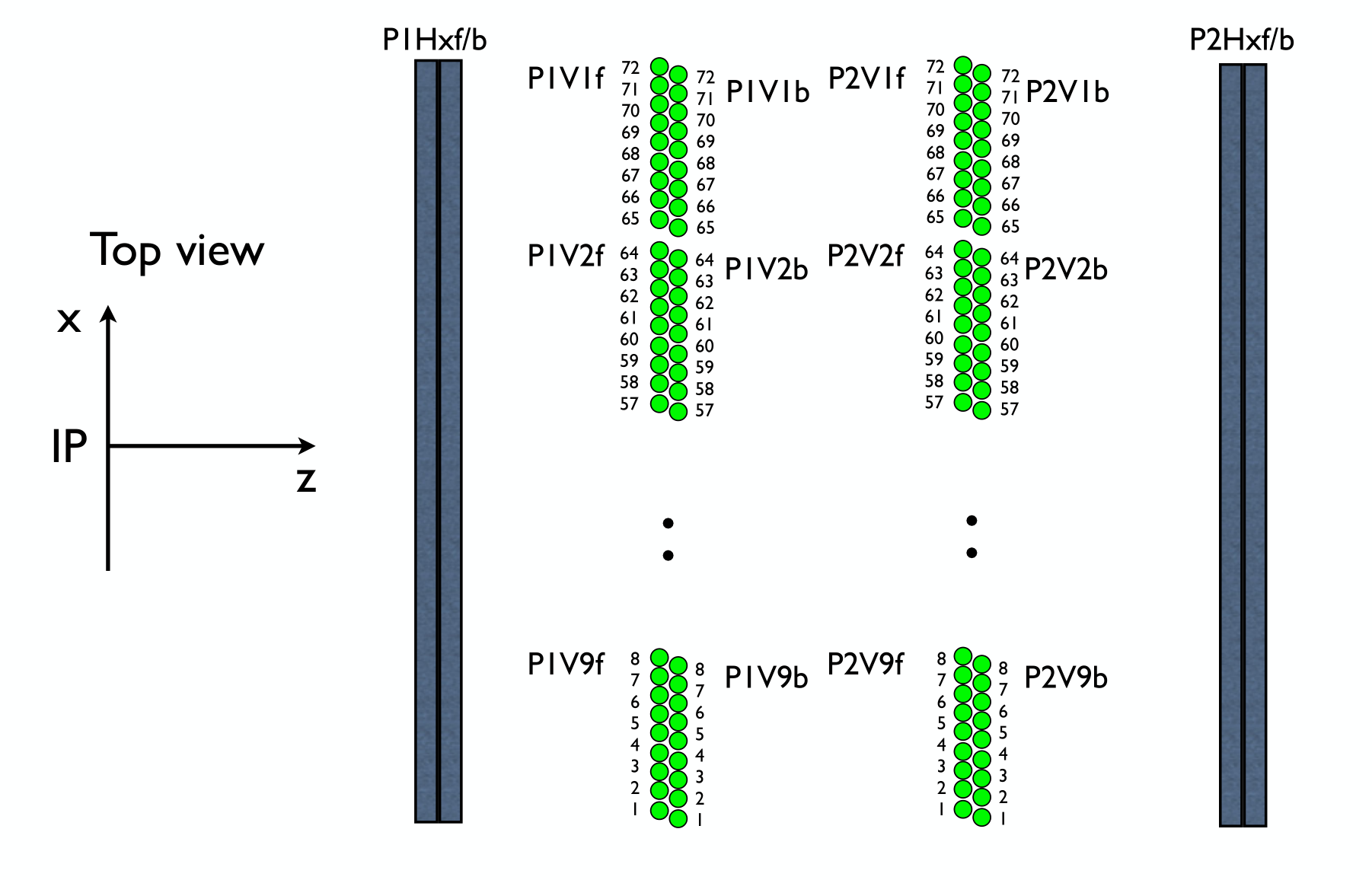}
   \\
    \includegraphics[width=\columnwidth]{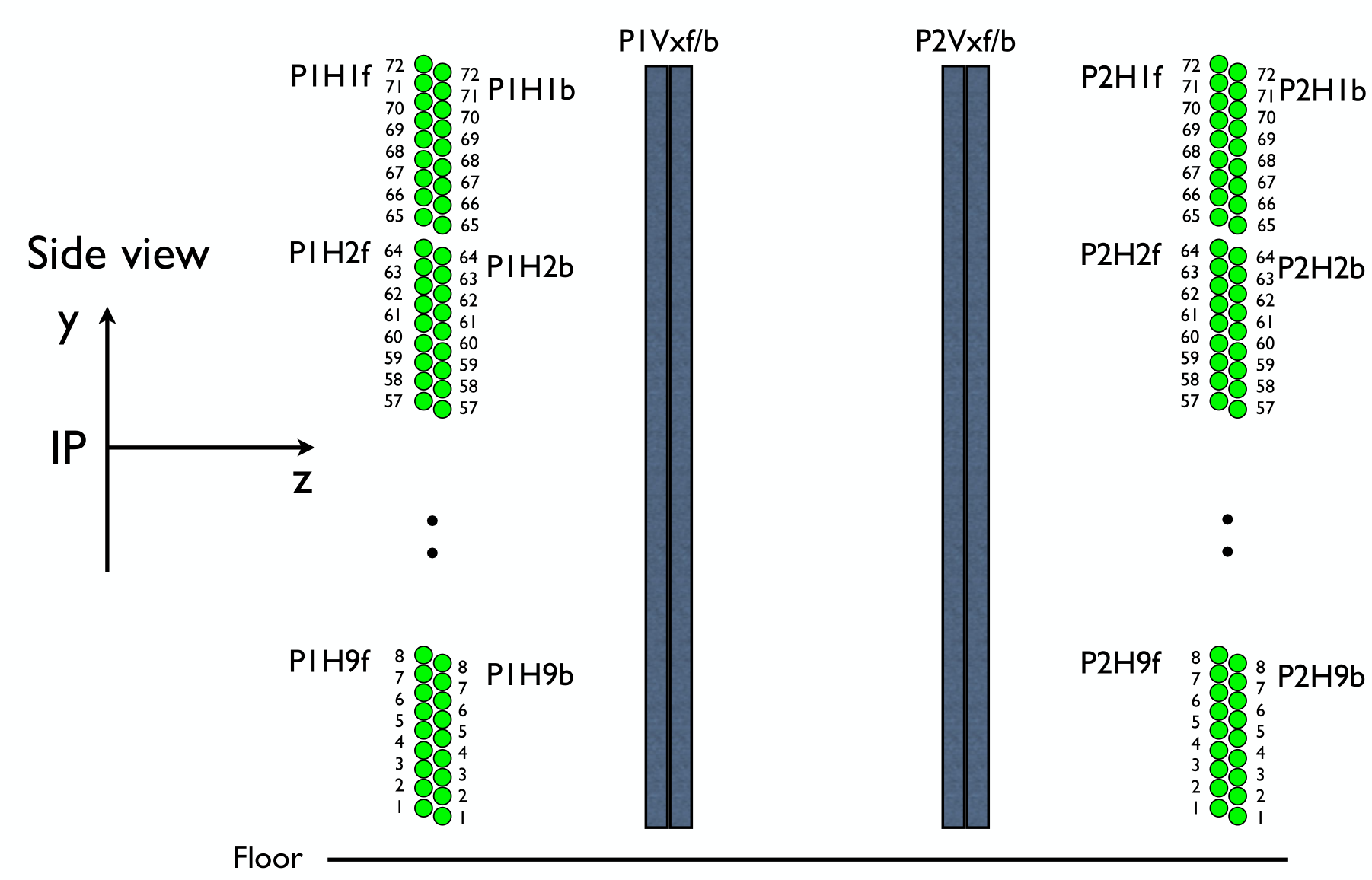}
    \caption{(Upper) Top (x-z) view of the proportional tube layout.  (Lower) Side (y-z) view of the proportional tube layout. \label{fig:proptube:view}}
\end{figure}

A typical high energy muon traverses two proportional tubes in each plane and induces hit signals on two anode wires.  Groups of 16 proportional tube anodes are read out independently through Nanometric Systems N-277 16-channel Amplifier/Discriminator cards with a common threshold preset externally\footnote{Nanometric Systems Inc., Oak Park, IL.  The N-277 data sheet is available at \url{https://hallcweb.jlab.org/experiments/hks/datasheets/nanometric.pdf}.}. 

The small, momentum-dependent deflection of the track angles before and after the iron absorber is used as the key signature of a muon track.  The position resolution has been studied with cosmic rays and with $J/\psi$ dimuon samples.  The residual distributions shown in Fig.~\ref{fig:res_proptube} were measured under the same condition as those of the drift chambers.  The asymmetric shapes of these distributions arise from inaccuracies of the relation between drift time and drift distance, particularly from tube-by-tube deviations.  The distribution widths are about 0.5 mm, which is more than sufficient for muon identification purposes. For the muon identification, 8 hits from 4 planes of the proportional tubes are used to form a muon road pointing back to the target. With a maximum drift time of 650 ns, the proportional tube can have acceptable performance with a singles rate up to 2 MHz/wire.  In normal operation the hit rate was typically below 1 MHz/wire.
\begin{figure*}[tb]
  \begin{center}
    \includegraphics{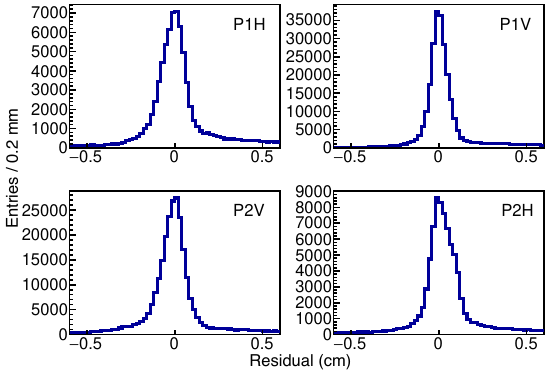}
    \caption{The residual distributions from the individual layers of the proportional tubes. \label{fig:res_proptube}}
  \end{center}
\end{figure*}
data set\section{Trigger \label{sec:trigger}}

The SeaQuest trigger uses discriminated signals from the hodoscope counters and is designed to be sufficiently flexible to quickly accommodate changes in the spectrometer, beam conditions, and physics goals. The trigger is optimized to accept high-mass (\unit[4-10]{GeV/c$^2$}) dimuons originating from the targets. In order to keep the trigger rate low enough to maintain an acceptable data acquisition (DAQ) dead time, most other sources of dimuons, such as $J/\psi$ decays, are suppressed.  A detailed description of the main components of the firmware design (the TDC, the delay adjustment, and the trigger logic) is given in Ref.~\cite{Shiu201582}. Details about the final design and performance of the trigger system, including the trigger logic optimization, can be found in Ref.~\cite{McClellanThesis}

\subsection{Overall Structure}
The SeaQuest Trigger System uses 9 CAEN V1495 VME modules that include an Altera EP1C20F400C6 FPGA, and a ``Trigger Supervisor'' VME module designed at Jefferson Laboratory~\cite{TS}. The arrangement of these modules is depicted in Fig.~\ref{fig:trigger_schematic}. The trigger consists of three separate ``Levels'' of V1495 modules. Level 0 contains four V1495 modules, one for each hodoscope ``quadrant'' (upper bend plane, lower bend plane, upper non\-bend plane, and lower non\-bend plane). Level 0 operates in two distinct modes. In ``Production'' mode (normal data-taking), Level 0 simply passes the input signals from the hodoscopes directly through to the Level~1 modules. In ``Pulser'' mode, Level 0 generates arbitrary hit patterns on its outputs. Level 0 Pulser Tests are used to verify the behavior of Level~1 and Level~2. Level~1 also consists of four V1495 modules, each taking output signals from one Level 0 board as input. Level~1 is responsible for finding four-hit track candidates in each quadrant. The track candidates are grouped into bins, and these bins are sent to Level~2 as a bit string (up to 32 bits from each Level~1 module).  Level~2 is a single board, which takes up to 32 channels of input from each Level~1 board. Level~2 is the ``track correlator.'' Level~2 forms all possible pairs of track candidates from Level~1, applies firmware-defined selection criteria, and sends five independent  output triggers to the Trigger Supervisor. (See Sec.~\ref{sec:modules}.)

\begin{figure}[tb]
\centering
\includegraphics[width=\columnwidth]{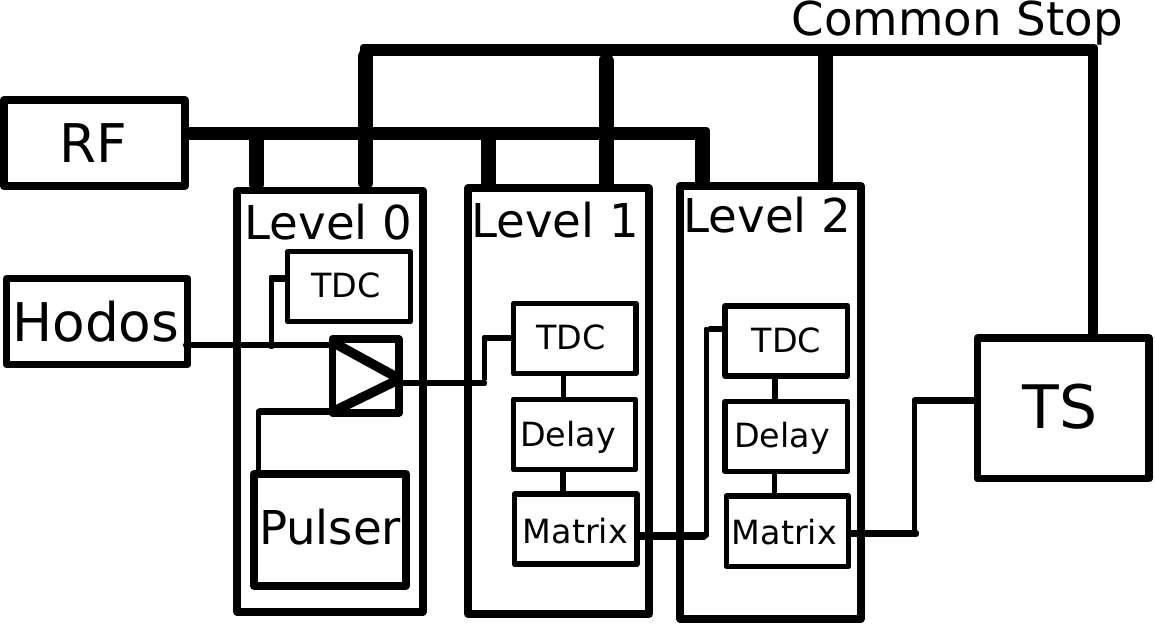}
\caption{Trigger hardware schematic~\cite{McClellanThesis}.}
\label{fig:trigger_schematic}
\end{figure}

\subsection{Firmware}
Custom firmware was written to meet the requirements for the SeaQuest trigger.   The Level~1 and 2 firmwares are largely identical, differing only in the content of the logic pipeline. There are three main parts of the firmware: the TDC block, the delay adjustment pipeline, and the trigger matrix. Level 0 shares the TDC block and the delay adjustment pipeline, but the trigger matrix is unused. Additionally, Level 0 contains the Pulser-mode firmware components.

The TDC digitizes the input signals within each V1495 module. A four-phase sampling scheme is used to achieve 1.177 ns resolution with a \unit[212.4]{MHz} clock. An onboard PLL (Phase Locked Loop) is used, with a 16/4 ratio, to generate the \unit[212.4]{MHz} clock from the external \unit[53.1]{MHz} MI (Main Injector) RF clock. The MI RF clock is synchronized with the RF structure of the delivered proton beam. The TDC block generates four clocks from the fast clock, with phases offset by $90^\circ$ ($0^\circ$, $90^\circ$, $180^\circ$, $270^\circ$). The input of each channel is sampled with all four clocks, to achieve an effective TDC resolution of \unit[1.177]{ns}.  The trigger is not able to distinguish between accidental hits and real hits when it compares with the programmed trigger roads.  All hodoscope hits are treated equally.

The delay-adjustment pipeline aligns the timing of the input signals and provides event storage. It consists of RAM blocks where digitized hits are stored in 16 clock-tick (18.8 ns) bins. The input timing for each channel is individually adjustable in \unit[1.177]{ns}  steps. Each 16 clock-tick bin (within each channel), can hold only one hit. If multiple hits arrive in the same TDC bin, only the latest hit is stored. This pipeline also serves as event storage for DAQ readout of the V1495 TDC contents. 

The trigger matrix is a lookup-table-based trigger logic implementation. The digitized hits from the TDC block are sent to the Trigger Matrix where  they are combined to generate the output signals. For each RF bin, the Level~1 Trigger Matrix compares the pattern of hits against a list of hit patterns designated as ``Trigger Roads.'' Any and all matching patterns generate an output bit. The output bits are binned by charge and average $x$-momentum ($p_x$) of the track. There are twelve, \unit[0.5]{GeV/c}-wide $p_x$-bins for each charge, so each Level~1 board outputs a 24-bit word for each RF clock cycle. If multiple patterns match a single charge/$p_x$ bin, the output bit is still set to True. The Level~2 Trigger Matrix checks all possible combinations of individual roads (found by Level~1) against a lookup table of  valid ``di-roads.'' The v1495 trigger system is veto-free. If the hits from a single RF cycle can satisfy a valid pair of roads, the trigger will fire, regardless of presence additional `accidental' hits.  Although possible in principle, the Level~2 trigger does not utilize the non-bend plane information nor the $p_x$-bin information. The definitions of the Level~2 output triggers are given in Tab.~\ref{tab:L2trigger}.

\begin{table*}
\centering
\caption{The five outputs of the Level~2 trigger module. ``Matrix 1'' is the main production trigger. ``Matrix 3'' is used to estimate combinatoric background contributions. The other three are not currently used in the analysis.  The column labeled ``Side'' denotes the combination of either the top or bottom (T or B respectively) that the triggering tracks were found.\label{tab:L2trigger}}
\begin{tabular}{lllll}
 \\ \hline \hline
 Name & Side & Charge & $p_x$ Req.& Notes \\ \hline
 Matrix 1 & TB/BT & $+-/-+$ & None & Main physics trigger \\
 Matrix 2 & TT/BB & $+-/-+$ & None & Same-Side trigger \\
 Matrix 3 & TB/BT & $++/--$ & None & Like-Charge trigger \\
 Matrix 4 & T/B & $+/-$ & None & All singles trigger \\
 Matrix 5 & T/B & $+/-$ & $p_x > $\unit[3]{GeV/c} & High-$p_T$ singles trigger \\ \hline \hline
\end{tabular}
\end{table*}

Level 0's ``Pulser Mode'' requires additional firmware blocks for storing hit patterns and generating ``pulser'' output signals. In ``Pulser Test'' mode, Level 0 reads hit patterns from text files, loads those hit patterns into RAM blocks, and generates output signals based on those patterns. With Level~1 and Level~2 operating normally, the ``pulser'' output from Level 0 is treated identically to real hodoscope signals. Comparing the output of Level~2 with the expectation based on the loaded pattern, the behavior of the Level~1 and Level~2 trigger logic can be verified.

Over the life of the experiment, the trigger pattern was reconfigured in response to changes in experimental conditions, test data needs and spectrometer configurations.  Downloading a new trigger pattern to the V1495 system could be accomplished in less five minutes.  In general, the DAQ system was stopped while a new trigger pattern was downloaded.  ``Pulser Mode'' tests required additional time and were completed when the experiment was not scheduled to be recording data for other reasons.

\subsection{Performance} 
SeaQuest's FPGA-based trigger system has performed, and continues to perform, quite well throughout data-recording. Together with the Beam Intensity Monitor, the trigger is able to preferentially select candidate dimuon events  out of a very high-rate environment. Continuous improvements to the trigger system have led to improved signal acceptance, background rejection, and self-monitoring.

The trigger acceptance in mass, relative to the geometric acceptance of the hodoscopes is shown in Fig.~\ref{fig:trigAcc} (top) for the  data set 2 and data set 3 ``matrices.'' The trigger roads are chosen to significantly suppress events with $M < $\unit[$4$]{GeV/c$^2$}.  Dimuons generated by charmonium decays would otherwise dominate the event rate and overwhelm the DAQ. Above \unit[4.2]{GeV/c$^2$}, the sample of reconstructed dimuons is dominated by the Drell-Yan process. Since the main physics goals of SeaQuest cover a broad range in $x_2$, the trigger acceptance must be relatively flat  in $x_2$. Figure~\ref{fig:trigAcc} (bottom) shows the trigger acceptance relative to the geometric hodoscope acceptance for events with mass greater than \unit[4.2]{GeV/c$^2$} versus $x_2$ as defined in leading order.  Next-to-leading order terms will modify this by less than 10\%.

\begin{figure}[tb]
\centering
\includegraphics[width=\columnwidth]{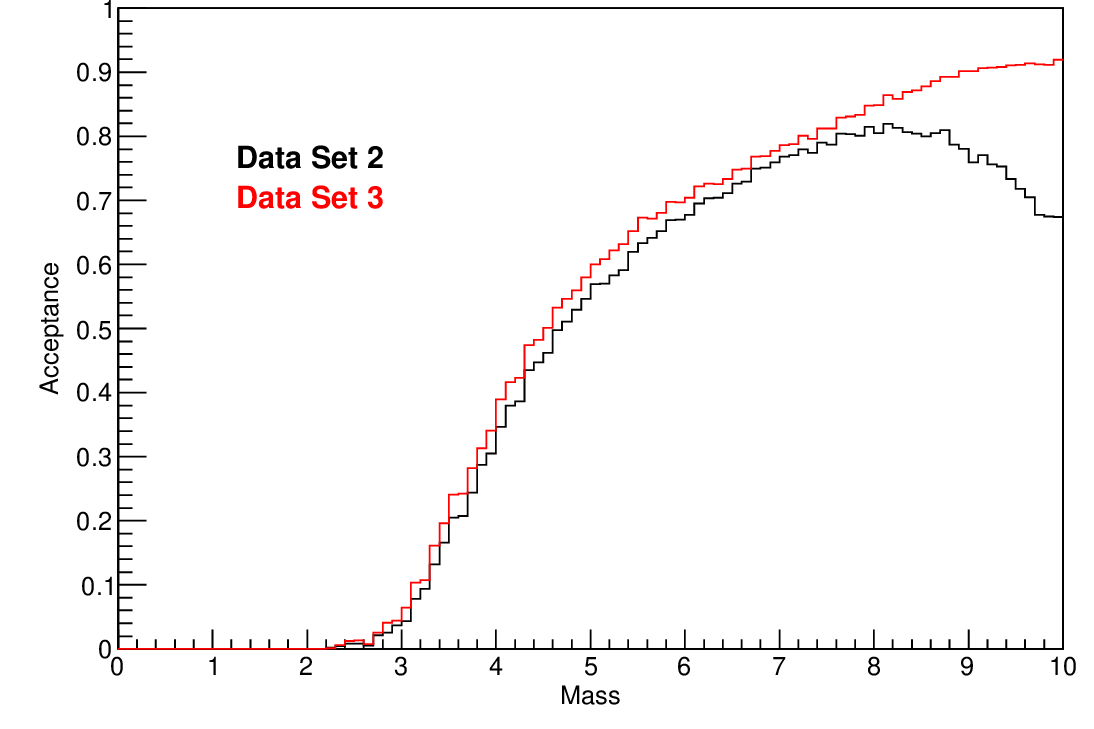}
\\
\includegraphics[width=\columnwidth]{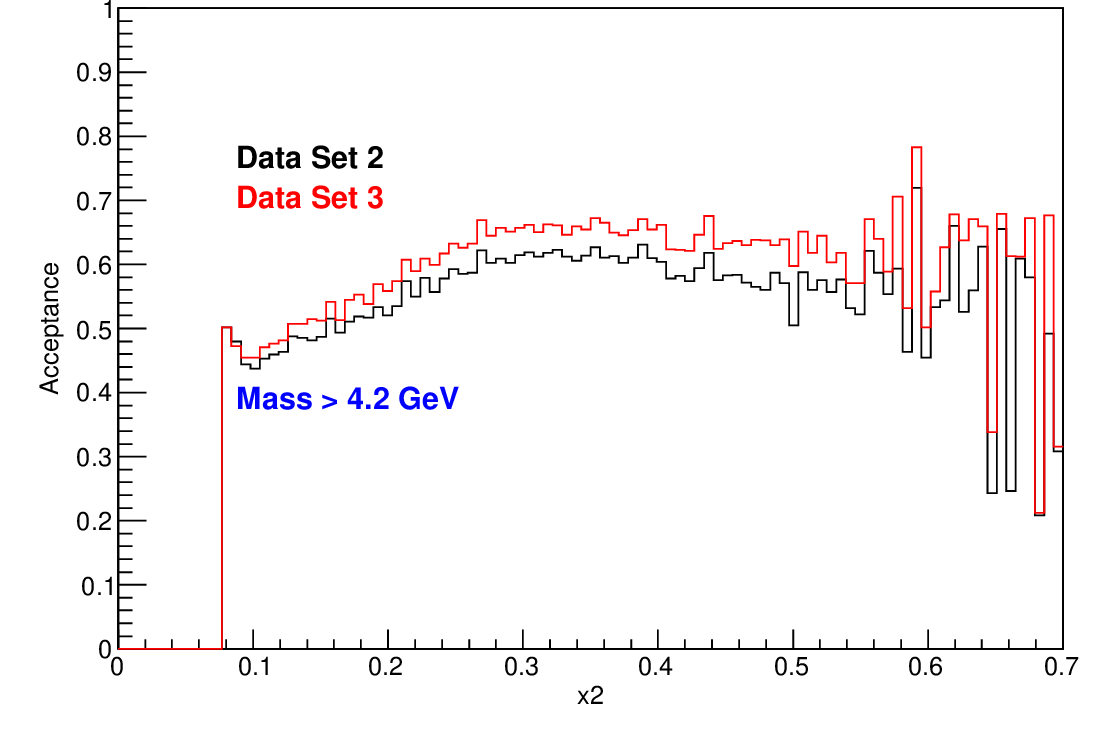}
\caption{Trigger Acceptance versus mass (top) and $x_2$ (bottom) for data sets 2 and 3 trigger matrices as defined in leading order.  Next-to-leading order terms will modify this by less than 10\%.~\cite{McClellanThesis}. \label{fig:trigAcc}}
\end{figure}

The consistent and reliable behavior of the V1495 trigger system was verified through extensive \textit{in situ} and bench pulser testing. The analysis of recorded data verified the internal consistency of the trigger in real experimental conditions. That is, by comparing the recorded input hits to the generated trigger output, the trigger system is found to agree with expectations. The V1495 trigger system, as well as the entire DAQ system, was located downstream of FMag and the radiation shielding around the target system, and radiation did not appear to affect the performance of the V1495 system.
data set\section{Data Acquisition Systems\label{sec:daq}}

Data acquisition for SeaQuest is divided into three separate systems based on timing and bandwidth requirements that could not be easily met with a single central system.  The three subsystems are called ``Event DAQ,'' ``Scaler DAQ,'' and ``Beam DAQ.'' The Event DAQ records the event-by-event main detector information and the trigger timing. The Scaler DAQ records the scaler information on a 7.5 kHz clock and at the end of the spill. The Beam DAQ records information from the beam line Cerenkov detector (discussed in Sec.~\ref{sec:BIM}).   Both the Event and Scaler DAQs use the VME-based ``CODA'' (CEBAF On-line Data Acquisition)~\cite{coda} system developed by the Thomas Jefferson National Accelerator Facility (JLab).  The VME modules used by SeaQuest are described in Sec.~\ref{sec:modules}.

\subsection{Event DAQ\label{sec:eventdaq}} 
The Event DAQ digitizes the signals from the spectrometer on an event-by-event basis.  This system uses multiple front-end VME crates that operate in parallel, each reading a specific part of the spectrometer.   Each VME crate contains a single board processor, a trigger interface (or supervisor) and a number of TDCs or other modules to be read out.  In SeaQuest, the number of front-end crates varied with the spectrometer configuration, but in general, there were about 14 front-end VME crates along with one trigger supervisor (TS) crate. While the details of the CODA framework can be found elsewhere~\cite{coda}, some of the basic features are described here.    The data from the individual front-end VME crates is transmitted over gigabit Ethernet using a private network.

The system is triggered by either a signal from the trigger matrix (see Sec.~\ref{sec:trigger}) or from NIM diagnostics  triggers.  These signals are processed by the TS~\cite{TS}.  The TS can receive up to 12 different input triggers. The first four triggers can be pre-scaled by up to 24 bits and are used for the matrix triggers from the CAEN V1495 logic. The second four triggers (MATRIX5, NIM1-3) can be pre-scaled up to 16 bits. The remaining four triggers are not pre-scalable. When the TS accepts a trigger, it transmits a signal to the trigger interface (TI) in each front-end VME crate.   The TI acknowledges the trigger and alerts the VME processor to read out the event.  The TS also generates the signal that is used for the TDC common stop.  When each front-end VME crate has completed its readout, the TI signals the TS that the read out is complete.  The TS also sends a trigger to QIE (see Sec.~\ref{sec:BIM}).  The QIE system retains data from a 12 to 16 RF bucket time window around the trigger to measure the beam intensity before and after each trigger. This output is encoded and read by a scaler-latch located in one of the front-end VME crates. 

During the recording of data sets 1-6, readout time was approximately \unit[150]{\us}. Since most of the dead time was associated with data transfer over the VME backplanes, a buffered readout scheme was developed in which data is stored locally in each TDC module during the four-second slow spill and transferred over the VME backplanes only between spills. This reduced the readout time to approximately \unit[30]{\us} during the spill.  This was deployed in fall 2016 between data sets 6 and 7.

\subsection{DAQ Electronics\label{sec:modules}}

Each VME crate is equipped with a single board computer, or Read Out Controller (ROC). Each processor is responsible for the readout of the VME modules within its crate. The SeaQuest systems use  Motorola MVME55000 and MVME61000 boards running the VXWorks operating system, and Concurrent Technologies VX913/012-13 running Linux.

Trigger inputs come from the TS that accepts trigger inputs from the V1495 system and from NIM logic. It prescales some trigger inputs and fans out the trigger to the rest of the Event DAQ. The TI in each VME crate receives the trigger and passes it to the ROC in that VME crate. When the ROC is ready for another trigger, it alerts the TI. The TI in turn sends a signal to the TS indicating it is ready for another trigger. The TS inhibits  triggers until all TIs are ready. A Struck SIS3610 module provides the handshake between the NIM trigger logic and Scaler DAQ and sends the interrupt to the scaler ROC when the scaler DAQ accepts the trigger~\cite{SIS3610}.

The discriminated signals from the hodoscopes, drift chambers and proportional tubes are converted into time signals by custom Time-to-Digital Converter (TDC) modules~\cite{wang2014field}. These modules have 64-channel input in a 6U VMEbus form factor and are equipped with a low-power and radiation-hardened Microsemi ProA-SIC3 Flash based FPGA~\cite{ProASIC3}. The firmware digitizes multiple input hits of both polarities while allowing users to turn on a multiple-hit elimination logic to remove after-pulses in the wire chambers and proportional tubes. A scaler is implemented in the firmware to record the number of hits in each channel.

The Scaler-Latch (SL) uses the same hardware as the TDCs. The only  difference is the firmware. The SL receives a 16 channel LVDS output from the Beam Intensity Monitor interface module (see Sec.~\ref{sec:BIM}). The SL is initialized when all 16 channels are low for longer than \unit[12.8]{$\mu$s}. The SL is triggered when it receives a high signal and then every 100 ns, the SL records 16 bits data until 128 words are captured for each trigger. If  there are more than 128 words, the extra words will start a new trigger and overwrite the buffer.

The VME Scaler used for this system is a custom made 32 bit per channel 140 MHz scaler produced by IPN Orsay for the G0 Experiment at JLab~\cite{Scale32}.

\subsection{Scaler DAQ} 

The Scaler DAQ system operates independently of the Event DAQ.  This system is alive whether or not the Event DAQ is recording events and allows the experiment to monitor the spectrometer, trigger and beam conditions.  The system is comprised of one VME crate reading out four scalers.   One of these scalers is triggered by the coincidence of a \unit[7.5]{kHz} gate generator and the beam spill signal.  This records the \unit[7.5]{kHz} response of two unrelated hodoscopes and can be used to calculate the duty factor of the incoming beam.   The other three scalers are triggered by the BOS or EOS signals and record spill-level rates. Data collected by these spill-level scalers include the number of times each Event DAQ trigger is satisfied, intensity of the beam, and the rates of the hodoscope arrays. As with the Event DAQ, the readout of the VME-based DAQ is done using CODA. An independent program analyzes the data in realtime to monitor the performance of the detector and trigger, as well as the quality of the beam. A particularly useful diagnostic was a fast Fourier transform of the beam intensity recorded at \unit[7.5]{kHz} during the 4 s spill, as shown in Fig.~\ref{fig:fft}.

\begin{figure}
  \begin{center}
    \includegraphics[width=\columnwidth]{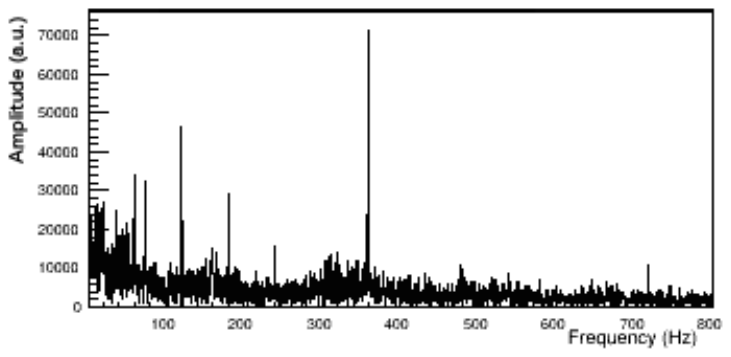}
    \caption{A fast Fourier transform of the beam intensity recorded at \unit[7.5]{kHz}.  The harmonics of the \unit[60]{Hz} line frequency are visible at 120, 360 and even \unit[720]{Hz}.   Similar plots were useful to the Fermilab Accelerator Division in diagnosing AC ripple on power supplies. \label{fig:fft}}
  \end{center}
\end{figure}

\subsection{Beam DAQ} 

The Beam DAQ reads the data from a Cerenkov detector in the proton beam (see Sec.~\ref{sec:BIM}).  This records the  \unit[53]{MHz} structure of the beam, i.e. the intensity of each RF bucket. Its calculation of the \unit[53]{MHz} duty factor $DF=\frac{<I>^{2}}{<I^{2}>}$ is the primary measure of the quality of the delivered beam that accelerator operators use for tuning. The \unit[53]{MHz} readout capability also enables a very accurate determination of live time.

There are four types of data that are recorded by the QIE board during the spill: (a) the intensity of each RF bucket; (b) the number of protons inhibited due to high instantaneous intensity for each inhibit generated; (c) the number of protons missed because the trigger system was busy during readout. This number excludes inhibited protons to avoid double counting; and (d) the sum of beam intensity, $I$, and intensity squared $I^2$ for the spill, from which the duty factor is calculated.

The Beam DAQ commences the readout of each of these blocks of data when the EOS signal is seen. The block of pulse height data for all buckets is about \unit[300]{MB}. To read this much data in time to analyze it and be ready for the next spill, the DAQ program uses Boost's implementation of multithreading\footnote{http://www.boost.org}. Three threads are used to read the data from the Beam Intensity Monitor board's three Ethernet chips, and up to eight threads are used to analyze the data expeditiously. The analyzed data is displayed on a public webpage so that shift personnel and accelerator operators can monitor the quality of the beam.  A representative plot of part of a spill's beam intensity is shown in Sec.~\ref{sec:beam}, Fig.~\ref{fig:cherenkovPlot}.

\subsection{Slow Controls}

A suite of auxiliary scripts runs on a gateway server to handle slow control, synchronization of DAQ data streams, and status monitoring. These scripts utilize the standard Experimental Physics and Industrial Control System (EPICS)~\cite{dalesio1994experimental} software package to communicate the values of process variables across various servers.

Since the three independent DAQ systems write output to three separate files, the decoding codes must know how to link to the same beam spill. This is accomplished by assigning a spill ID number for each spill and writing it into the data stream of each DAQ. One master spill ID is stored in a file, which is updated each spill. A script will increment the number stored in this file each time the EOS signal is seen. The script further writes this value into the easily accessible but volatile memory of the EPICS server. The file is inserted directly into the CODA output files of both the Event DAQ and Scaler DAQ, so that the spill ID is stored directly inside the data recorded during that spill. Programs performing realtime analysis read the spill ID via EPICS and similarly attach this number to any output they create.

Slow control data are collected when the EOS signal is delivered. The data describe the accelerator, target, and environmental conditions during the beam spill. The accelerator information describes the intensity and quality of the beam delivered, configuration of the accelerator, and status of SeaQuest's focusing and analysis magnets. These data are collected by Fermilab's accelerator controls network (ACNET)~\cite{nagy1986fermilab} and retrieved by SeaQuest scripts from the ACNET using its XML-RPC server. Target data are read from an EPICS instance that interfaces directly with the target system's open platform communication (OPC).  The target in the beam, target rotation pattern, and pertinent pressures and temperatures of the cryogenics are recorded. Environmental conditions are monitored by a multichannel digital multimeter that communicates over Ethernet and gathers data from temperature, pressure, and humidity sensors deployed throughout the detector hall.  The  values are reported via a slow control script on request that requires less than 10 seconds to run.  The temperatures measured include DAQ crates, and ambient temperatures at lower and upper parts of the detector hall.  The humidity sensors have thermistors so that dew points can be calculated.  Combined with pressure sensors these 
can be used to understand high voltage leakage currents.  These data are written to a file and inserted into the CODA output files of both the Event DAQ and Scaler DAQ. The spill ID is included in this file for redundancy, should the synchronization protocols described above fail.

The status of data collection is monitored in realtime to augment the ability of shifter personnel and experts to ensure high-quality data. Critical components monitored include the accelerator spill signals and whether beam is being delivered, that all three DAQ systems are alive and realtime monitoring is up-to-date, and whether there is adequate disk space. The results of these checks are output to a public webpage. If there is a problem that jeopardizes data collection, the shifter is notified with an audible alarm and the appropriate expert is alerted by text or e-mail.

\section{Summary and Future Measurements\label{sec:summary}}

The SeaQuest spectrometer described in this paper was designed to detect dimuon pairs produced in 120 GeV proton-proton and proton-nucleus collisions, with an emphasis on accepting dimuons produced in the decay of a high mass virtual photon or meson.  The spectrometer was modeled after earlier, highly-successful dimuon spectrometers at Fermilab.  The spectrometer has been in operation since 2012 collecting data for the SeaQuest measurement of $\dbar/\ubar$.  As a combined system, the spectrometer has a measured mass resolution of \unit[0.21]{GeV/c$^2$} at the $J/\psi$ mass as shown in Fig.~\ref{fig:massplot}, in excellent agreement with Monte Carlo simulations.   

\begin{figure}[tb]
  \centering
  \includegraphics[width=\columnwidth,clip]{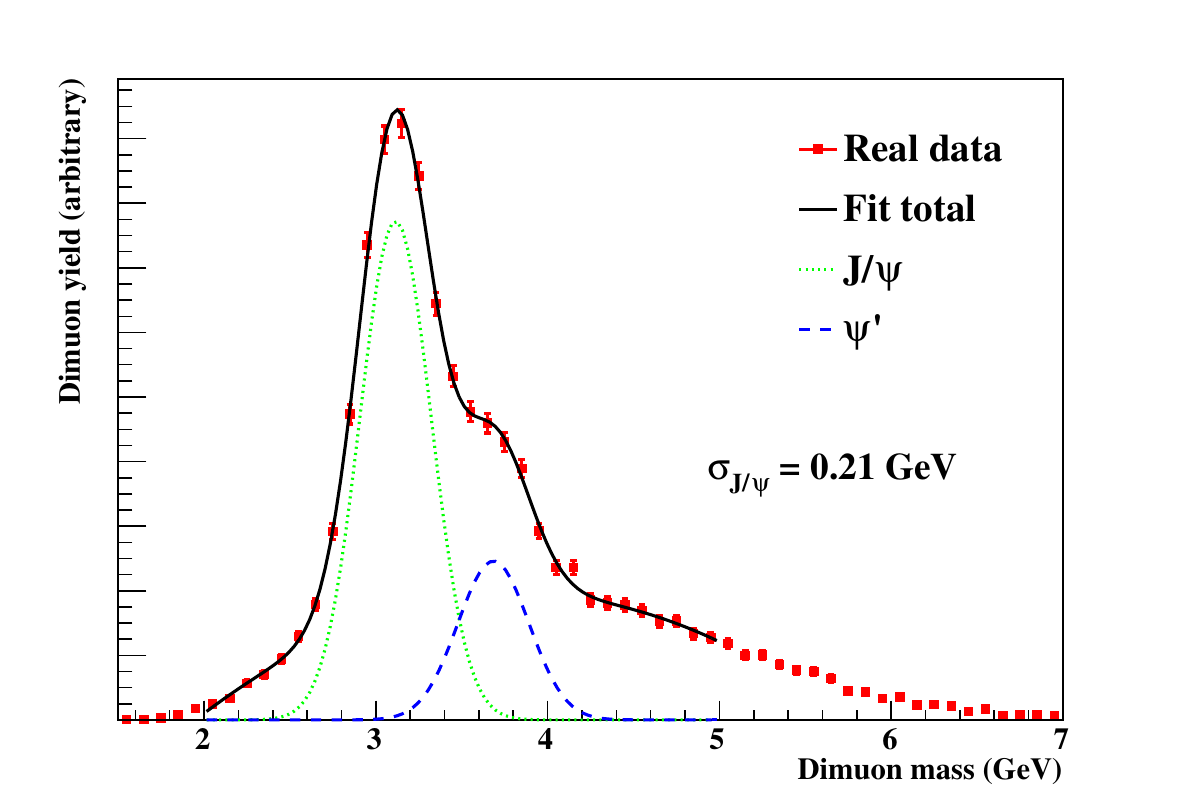}
  \caption{A fit to a subset of the measured mass spectra from the SeaQuest spectrometer with the hydrogen target.  The fit includes a Gaussian distribution for $J/\psi$ and $\psi^\prime$, in combination with a polynomial.  The resolution is \unit[0.21]{GeV/c$^2$}, in good agreement with Monte Carlo simulations.    \label{fig:massplot} }
\end{figure} 

Since construction of the SeaQuest spectrometer for this measurement began, additional measurements have been proposed using the spectrometer either parasitically or in dedicated runs, including (a) a search for dark sector photons~\cite{Gardner:2015wea}; (b) a study of the transverse polarization of sea quarks by measuring the Sivers asymmetry with a transversely polarized target~\cite{LorenzonPolarized}; and (c) a high statistics test of the Sivers DIS/Drell-Yan sign change with a polarized proton beam~\cite{LorenzonPolarized}.  Funding is being sought for the  polarized measurements, which have received Stage I approval from Fermilab.  The dark photon search is running parasitically with the current measurement.
\section{Acknowledgements}

We wish to thank Sten Hansen and Terry Kiper of the Electrical Engineering Department of Fermilab's Particle Physics Division.  Sten designed much of the SeaQuest electronics, including the Beam Intensity Monitor Readout module, the wire chamber front-end system (ASDQ cards and Level Shifter Boards), and high rate phototube voltage dividers for the station 1 and 2 hodoscopes.  Terry wrote all associated micro controller code and has maintained these systems.  We would also like to thank the JLab CODA group in particular David Abbot and Ed Jastrzembski for their help with CODA installation.

This work was supported in part by US Department of Energy grants DE-AC02-06CH11357, DE-FG02-07ER41528, DE-SC0006963;  US National Science Foundation under grants PHY 0969239, PHY 1452636, PHY 1505458; the DP\&A and ORED at Mississippi State University; the JSPS KAKENHI Grant Numbers 21244028, 25247037, 25800133; Tokyo Tech Global COE Program; Yamada Science Foundation of Japan; and the Ministry of Science and Technology (MOST), Taiwan.  Fermilab is operated by Fermi Research Alliance, LLC under Contract No. DE-AC02-07CH11359 with the United States Department of Energy.
\end{twocolumn}

\onecolumn

\vspace*{0.25in}\hrule\vspace*{0.25in}

\begin{multicols}{2}
\section*{\refname}
\addcontentsline{toc}{section}{\refname}

  \bibliographystyle{elsarticle-num}
  \bibliography{SeaQuestNIM}
\end{multicols}
  \end{document}